\documentclass[a4paper,12pt]{article}
\usepackage{jheppub}
\usepackage[dvipsnames]{xcolor}
\usepackage{tikz}
\usepackage{pgfplots}
\usepackage[T1]{fontenc}
\usepackage[]{slashed}
\usepackage[]{bm}
\usepackage{physics}
\usepackage{dsfont}

\usepackage[force]{feynmp-auto}

\usepackage{hyperref}
\hypersetup{colorlinks=true,linkcolor=magenta,anchorcolor=green,citecolor=cyan,filecolor=black,menucolor=black,urlcolor=brown}
\makeatletter

\usepackage{lipsum}

\DeclareMathOperator\arcsinh{arcsinh}
\DeclareMathOperator\arccosh{arccosh}

\DeclareMathOperator\Li{Li}

\usepackage{graphicx}
\usepackage{epstopdf}
\usepackage{bbm,amsmath,graphicx,amssymb,amsfonts,amsthm}

\def\sq[#1,#2]{\left[#1\,#2\right]}
\def\an[#1,#2]{\left\langle#1\,#2\right\rangle}
\def\spab[#1,#2,#3]{\left\langle#1|#2|#3\right]}

\def\spb(#1,#2){\left[#1\,#2\right]}
\def\spa(#1,#2){\left\langle#1\,#2\right\rangle}
\def\spakb(#1,#2,#3,#4,#5){\left\langle#1|\slash \!\!\!\;\!\! #2\slash  \!\!\!\;\!\! #3 \slash  \!\!\!\;\!\!#4 |#5\right]}
\def\spbka(#1,#2,#3,#4,#5){\left\langle#1|\slash  \!\!\!\;\!\! #4\slash  \!\!\!\;\!\! #3 \slash  \!\!\!\;\!\!#2 |#5\right]}
\def\spaa(#1,#2,#3,#4){\left\langle#1|\slash  \!\!\!\;\!\! #2\slash  \!\!\!\;\!\! #3  |#4\right\rangle}
\def\spab(#1,#2,#3){(\ell_#1\cdot #2)}
\def\spakkb(#1,#2,#3){\left\langle#1|\slash \!\!\!\;\!\! #2 |#3\right]}
\def\spakkkb(#1,#2,#3){(#1\cdot #2)}
\def\spbkka(#1,#2,#3){\left[#1|\slash \!\!\!\;\!\! #2 |#3\right\rangle}
\def\spbkkka(#1,#2,#3){(#1\cdot #2)}
\def\spbb(#1,#2,#3,#4){\left[#1|\slash  \!\!\!\;\!\! #2\slash  \!\!\!\;\!\! #3  |#4\right]}
\def\Ttrma(#1,#2,#3,#4){{\rm tr}_{-}[\slash \!\!\!\;\!\! #1\slash  \!\!\!\;\!\! #2 \slash  \!\!\!\;\!\!#3\slash  \!\!\!\;\!\!#4]}
\def\Ttrmb(#1,#2,#3,#4,#5,#6){{\rm tr}_{-}[\slash \!\!\!\;\!\! #1\slash  \!\!\!\;\!\! #2 \slash  \!\!\!\;\!\!#3\slash  \!\!\!\;\!\!#4\slash  \!\!\!\;\!\!#5\slash  \!\!\!\;\!\!#6]}
\def\Ttrmc(#1,#2,#3,#4,#5,#6,#7,#8){{\rm tr}_{-}[\slash \!\!\!\;\!\! #1\slash  \!\!\!\;\!\! #2 \slash  \!\!\!\;\!\!#3\slash  \!\!\!\;\!\!#4\slash  
\!\!\!\;\!\!#5\slash  \!\!\!\;\!\!#6\slash  \!\!\!\;\!\!#7\slash  \!\!\!\;\!\!#8]}
\def\Dp(#1,#2){(#1\cdot #2)}

\def\dBox{\Box\kern-.1em\Box}
\def\dNPBoxs{\scalebox{.9}{$\bowtie$}\kern-.1em\Box}
\def\dNPBoxu{\Box\kern-.1em\scalebox{.9}{$\bowtie$}}

\def\beq{\begin{equation}}
\def\eeq{\end{equation}}
\def\bes{\begin{split}}
\def\ees{\end{split}}
\def\beqa{\begin{eqnarray}}
\def\eeqa{\end{eqnarray}}

\def\eeqa{\end{eqnarray}}
\def\ek[#1,#2]{(\varepsilon_{#1}\cdot k_{#2})}
\def\e[#1,#2]{(\varepsilon_{#1}\cdot \varepsilon_{#2})}
\def\s(#1,#2){{(\ell_#1\cdot\ell_#2)}}

\def\e{\epsilon}

\pgfdeclarelayer{bg}    
\pgfsetlayers{bg,main}  

\definecolor{Mathematica}{HTML}{ed192d}

\usepackage{float}

\usetikzlibrary{shapes.misc}
\usetikzlibrary{arrows}
\usetikzlibrary{decorations.markings}
\usetikzlibrary{positioning,fit}
\usetikzlibrary{patterns}

\tikzset{cross/.style={cross out, draw=black, minimum size=2*(#1-\pgflinewidth), inner sep=0pt, outer sep=0pt},
	cross/.default={2pt}}

 \preprint{\vbox{\hbox{\hphantom{XXXX}IPhT-t21/015}\hbox{\hphantom{X}CERN-TH-2021-052}}}
      
\title{Classical Gravity from Loop Amplitudes}

\author[a]{N. Emil J. Bjerrum-Bohr}
\author[a,e]{\!\!, Poul H. Damgaard}
\author[b]{\!\!, Ludovic Plant\'e}
\author[c,d,e]{\!\!, Pierre Vanhove}

\affiliation[a]{Niels Bohr International Academy, Niels Bohr Institute, University of Copenhagen, Blegdamsvej 17, DK-2100 Copenhagen, Denmark}
\affiliation[b]{ 20 rue Marie et Pierre Curie
92110 Clichy, France}
\affiliation[c]{Institut de Physique Theorique, Universit\'e Paris-Saclay,
CEA, CNRS, F-91191 Gif-sur-Yvette Cedex, France}
\affiliation[d]{National Research University Higher School of
  Economics, Russian Federation}
\affiliation[e]{Theoretical Physics Department, CERN, 1211 Geneva 23, Switzerland}

\abstract{We describe an efficient method for extracting the parts of
 $D$-dimensional loop integrals that are needed to derive observables in
 classical general relativity from scattering amplitudes. Our approach simplifies the soft-region method of integration by
 judiciously combining terms before the final integrations. We
 demonstrate the method by computing the required integrals for black-hole scattering 
 to the second Post-Minkowskian order in Einstein gravity coupled to scalars. 
We also confirm
 recent results at the third Post-Minkowskian order regarding
 universality and high-energy behavior of gravitational interactions
 in maximal supergravity.}

\begin{document} 
\maketitle
\flushbottom
\section{Introduction}\label{sec:intro}
The detection of gravitational waves from binary mergers~\cite{Abbott:2016blz,TheLIGOScientific:2017qsa}
has opened a new and exciting
avenue for testing Einstein's theory of gravity at extreme energies
and at relativistic velocities. Although detailed comparisons of gravitational wave signals eventually require
numerical general relativity, there is an urgent need for highly precise analytical computations as well since
these are what's fed into accurate catalogs of gravitational wave templates. A particularly promising new approach for
such analytical calculations is based on the so-called Post-Minkowskian expansion in general relativity
\cite{Damour:2016gwp,Damour:2017zjx,Bjerrum-Bohr:2018xdl,Cheung:2018wkq}. 
This is the special-relativistic regime of
general relativity where relative velocities of two gravitationally interacting bodies are not assumed to be much
smaller than the speed of light. As such, this situation is ideally suited for the relativistic quantum field theory
description of gravity when truncated to the classical sector. The $S$-matrix describing the gravitational scattering
of two massive objects from Minkowskian infinity to
Minkowskian infinity will indeed automatically include all orders in
velocity to any given order in  Newton's constant $G_N$. When
restricted to the classical part of scattering this should coincide with the Post-Minkowskian expansion of Einstein gravity.\\[5pt]
There are numerous subtleties associated with this new and promising approach that need to be understood.
A folk-theorem of relativistic quantum field theory states that the only basic observables are $S$-matrix elements
and this seems a perfect starting point for a scattering calculation for the classical limit. In practice, the
$S$-matrix needs to be defined carefully in the presence of massless fields and one could worry that such
quantum field theoretic subtleties could translate into effects that survive when taking the classical limit.
Another concern could be the application the field-theoretic method to gravity because of its coupling to the
energy-momentum tensor. Is the high-energy limit of gravitational scattering where Mandelstam variable $s$ is much bigger
than the sum of the two masses squared $(m_1 + m_2)^2$ then equal to massless scattering? At one-loop order, this
is the case~\cite{Bjerrum-Bohr:2018xdl}. At two-loop order, the calculations of~\cite{Bern:2019nnu,Bern:2019crd,Cheung:2020gyp,Kalin:2020fhe} does not lead
to that conclusion, and indeed the scattering angle from that two-loop order computation, when extrapolated to
very high energies, diverges. Recently, it has been shown~\cite{DiVecchia:2020ymx} that this disconcerting conclusion is resolved when
the classical parts of the scattering amplitude are extracted from the so-called soft region of the loop integrals,
rather than from what is known as the potential region, which is the natural starting point of a low-energy computation.\\[5pt]
This situation has called for a renewed focus on obtaining a more systematic framework in which to compute those parts
of gravitational loop diagrams that will contribute to classical scattering in general relativity. In this paper,
we will introduce a formalism that we will argue simplifies computations based on the soft-regions method~\cite{Parra-Martinez:2020dzs,DiVecchia:2020ymx,Herrmann:2021tct}. 
While the
purpose of the method is to identify those terms that contribute to classical scattering~\cite{Bjerrum-Bohr:2018xdl,Kosower:2018adc}, the formalism is also
immediately applicable to both the  pieces in the amplitude that are
more singular than the classical one in  the $\hbar \to 0$ limit and
to those that correspond to quantum mechanical corrections, which are
subleading in the $\hbar\to0$ limit. A main advantage of our method is that it significantly
reduces the number of master integrals that need to be computed. This is 
already an important simplification at two-loop order
and it will be of even more value at higher-loop  order.
Our independent calculation at two-loop order for ${\cal N} = 8$ supergravity confirms the recent results of~\cite{DiVecchia:2020ymx,DiVecchia:2021ndb,DiVecchia:2021bdo,Herrmann:2021tct}.\\[5pt]
Before going into the technical details of our calculation, we find it useful to highlight the origin of the additional classical terms of
the soft region as compared to those of the potential region. Intuitively, the Post-Minkowskian expansion can appear as
a natural resummation of terms of the Post-Newtonian expansion. In the free-particle massive propagator, very schematically,
and ignoring the split into energy and three-momentum,
\beq
\frac{1}{k^2 - m^2} ~\sim~ \frac{1}{k^2}\left(1 + \frac{m^2}{k^2} + \left(\frac{m^2}{k^2}\right)^2 + \cdots\right), \label{PN-expansion}
\eeq
and the potential-region integration indeed re-sums such contributions correctly at one-loop order. However, this
one-loop result is already quite non-trivial and it does not seem to carry over to higher orders. The reason is that inside the loop integrals,
what plays the role of mass $m$ will be combinations of momenta. At specific regions of the integrations what effectively acts
as a mass term in the propagator can therefore vanish. At these points, dimensional regularization, if used in conjunction with the above expansion,
will na\"\i vely put these terms to zero for the same reason that the massive tadpole
\begin{equation}\label{e:tadpole}
\int \frac{d^{4-2\epsilon}k}{(2\pi)^{4-2\epsilon}} \frac{1}{k^2 -
  m^2+i\varepsilon}=-i\pi^{2-\epsilon} {\Gamma(\epsilon-1)\over (m^2)^{\epsilon-1}},
\end{equation}
will be set, incorrectly, to zero in dimensional regularization if one
integrates order by order in the small-mass expansion of
eq.~(\ref{PN-expansion}). For  $m^2\sim \sigma^2-1$ such a tadpole
integral  has a $1/(\sigma^2-1)^\epsilon$ dependence which is the
origin of the radiation-reaction contributions. 
A particular consequence of the
non-commutativity between the integration and the small velocity
expansion results in the vanishing of the  master integrals evaluated
in section~\ref{sec:b3b5} if one integrates the term-by-term
expansion. Consequently, all the radiation-reaction contributions to
the two-loop classical contribution vanish.
\\[5pt]

Concretely, the classical part of the two-loop amplitude gets a
contribution from the master integral $\mathcal I_6(\sigma)$ which
evaluates to
\begin{equation}\begin{split}
\!\!\mathcal I_6(\sigma)=\frac{(4\pi e^{- \gamma_E})^{2\epsilon}
  \epsilon^3}{8\pi^3}&
\arcsinh\left(\sqrt{\frac{\sigma-1}{2}}\right)\!\!\left(\pi+2i\left(-1\over
    4  (\sigma^2-1)\right)^{\epsilon} 
  \arcsinh\left(\sqrt{\frac{\sigma-1}{2}}\right)\right)\\& \hspace{9.7cm}+\mathcal O(\epsilon^4),
\end{split}\end{equation}
where $\sigma=p_1\cdot p_2/(m_1m_2)$ with $p_i$ the momenta of the
incoming massive scalars of mass $m_i$, and $\epsilon=(4-D)/2$ is the
dimensional regularization parameter.   
 This integral, like all the master two-loop integrals,  has a regular static limit $\sigma\to1$ only for
$\epsilon$ negative as required for the infrared behavior (see for instance~\cite{Kosower:1999xi}).

Performing the series expansion in $\epsilon$ near 0 
(and negative) with  $\sigma$ greater than 1 fixed gives
\begin{equation}\label{eq:seriesepsilon}
\mathop{\textrm{series}}_{\epsilon \rightarrow 0^-} \mathcal
I_6(\sigma)= \frac{(4\pi e^{- \gamma_E})^{2\epsilon}
  \epsilon^3}{8\pi^3}
\arcsinh\left(\sqrt{\frac{\sigma-1}{2}}\right)\left(\pi+2i \arcsinh\left(\sqrt{\frac{\sigma-1}{2}}\right)\right)+\mathcal O(\epsilon^4),
\end{equation}
which matches the result in eq.~(D.3) of~\cite{Bern:2019crd}.
Whereas the series expansion first  in $\sigma$ near 1
(and greater than 1) with  $\epsilon$ small and negative fixed, gives
\begin{equation}\label{eq:seriesigma}
\mathop{\textrm{series}}_{\sigma \rightarrow 1^+} \mathcal
I_6(\sigma)
=\frac{(4\pi e^{- \gamma_E})^{2\epsilon} \epsilon^3}{8\pi^2} \arcsinh\left(\sqrt{\frac{\sigma-1}{2}}\right)+\mathcal O(\epsilon^4),
\end{equation}
which is the result in eq.~(4.111)
of~\cite{Parra-Martinez:2020dzs}. At this order in $\epsilon$ one
notices that the result in~\eqref{eq:seriesigma} does not have the
imaginary part present in~\eqref{eq:seriesepsilon}. This shows the
importance of keeping the contribution $(-1/(4(\sigma^2-1)))^\epsilon$
in order to obtain  the radiation-reaction contributions to the
classical part of the amplitude.  And because
$(-1)^\epsilon=1+i\pi\epsilon+\mathcal O(\epsilon^2)$ this factor contributes
to both the real and imaginary part of the classical terms of
the amplitude.  As we will show in section~\ref{sec:twoloop} this factor will be
instrumental in showing the relation  between the real
part and the infrared divergence of the imaginary part of the
radiation-reaction contribution at two-loop order~\cite{DiVecchia:2021ndb}.\\[5pt]
Of course, eventually, the result must be unambiguous, but the latter order of limits seems to require a different approach to determine correctly the integration constants of
the set of differential equations. We will employ the former order of limit here, and use it to clarify very precisely the difference between the recent results of two different computations in~\cite{DiVecchia:2020ymx,Parra-Martinez:2020dzs}. The computation of~\cite{Parra-Martinez:2020dzs} provides all the terms relevant for the amplitude in the
potential region. We should stress that our purpose here is to
focus only on how to extract all classical terms of scattering amplitudes at loop orders. \\[5pt]
At two-loop order the calculation very neatly separates so that the final 
answer for the amplitude is decomposed into all the terms from
the potential region plus new contributions that arise from the above mechanism. Some of these new terms at two-loop order are real and some are both imaginary
and divergent. These new terms were first identified in refs.~\cite{DiVecchia:2020ymx}, and we confirm those results in all details. A surprising feature is that the new terms which are real,
being of half-integer order when expanded at low energy,
find no place within the conventional Post-Newtonian expansion~\cite{Bini:2020nsb,Blumlein:2021txj} of the conservative part of the interaction Hamiltonian. These additional terms have been
identified as radiation-reaction pieces~\cite{DiVecchia:2020ymx,DiVecchia:2021ndb} that normally would have to be
treated separately in the Post-Newtonian expansion. A recent calculation of Damour~\cite{Damour:2020tta}  confirms this picture, also suggesting a short-cut towards identifying
the terms that must be added to the potential-region calculation in Einstein gravity, and which will render the high-energy scattering angle finite at this order. It is interesting to 
speculate how the appearance of these
new pieces may have repercussions
at the conventional level of the Post-Minkowskian expansion from three
loops and higher\footnote{We thank Enrico Herrmann, Julio
  Parra-Martinez, Michael Ruf, and Mao Zeng for clarifying comments on
  this.}. A first calculation of the fourth Post-Minkowskian result for the conservative part of
the interactions of scalar black holes has already appeared~\cite{Bern:2021dqo}.\\[5pt]
We outline the paper as follows. In section~\ref{sec:review} we 
review how to determine classical general relativity from scattering
amplitudes. We will discuss how it is possible to expand the scattering
amplitudes at second (in section~\ref{sec:2PMlevel}) and third  (in section~\ref{sec:3PMlevel})
Post-Minkowskian order in terms of a
set of classical basis integrals multiplied with coefficients
determined from unitarity. In section~\ref{sec:3PMlevel} we give the
results of the evaluation of the third Post-Minkowskian contribution
in maximal supergravity. We show that the classical contribution to
the amplitude is independent of the helicity configuration.
 We then compare the results  with the ones from the potential and soft
integration region.
In section~\ref{sec:evaluationdoubleboxes} 
we evaluate the double-box integral contributions to the maximal
supergravity amplitude. In section~\ref{sec:masterI} we determine the
basis of master integrals used in the evaluation of the two-loop
double-box integrals. In appendix~\ref{sec:deltatomaster} we explain
how to reduce the generic two-loop double-box integrals to the
two-loop master integral with a generalized propagator that we use in
this computation. Finally, we will conclude and look ahead. 
%
%
\section{General relativity from quantum field theory}\label{sec:review}
The Einstein-Hilbert Lagrangian minimally coupled to two massive scalar fields reads,
\begin{equation}
{\cal L}_{EH} = \int d^4 x \sqrt{-g} \Bigg[\frac{R}{16 \pi G_N}  + \frac12 g^{\mu\nu} ( \partial_\mu \phi_1\partial_\nu \phi_1 + \partial_\mu \phi_2\partial_\nu \phi_2)- m_1^2 \phi_1^2 - m_2^2 \phi_2^2\Bigg]\,,
\end{equation} 
in this equation $G_N$ denotes the Newton constant, $R$ defines the Ricci 
scalar and $g$ is the determinant of the metric: $g_{\mu\nu}(x)\equiv \eta_{\mu\nu} + \sqrt{32 \pi G_N}h_{\mu\nu}(x)$ expanded around a Minkowski background, ${\rm diag}\,\eta_{\mu\nu}\equiv(1,-1,-1,-1)$. \\[5pt]
We consider scattering events organised in a perturbative expansion,
\begin{equation}
{\mathcal M}(p_1,p_2,p_1',p_2')=\!\!\!\!\begin{gathered}
    \begin{fmffile}{Smatrix}
    \begin{fmfgraph*}(100,100)
\fmfstraight
\fmfleftn{i}{2}
\fmfrightn{o}{2}
\fmfrpolyn{smooth,filled=30}{e}{4}
\fmf{fermion,label=$p_1$,label.side=left}{i1,e1}
\fmf{fermion,label=$p_1'$}{e2,i2}
\fmf{fermion,label=$p_2'$}{e3,o2}
\fmf{fermion,label=$p_2$,label.side=right}{o1,e4}
\end{fmfgraph*}
\end{fmffile}
  \end{gathered}
\!\!\!\!=\sum_{n\,=\,0}^{ \infty} \mathcal{M}_{ L}(p_1,p_2,p_1',p_2') 
\,, \ \mathcal{M}_{ L} \sim {\cal O}(G_N^{L+1}),
\end{equation}
where $p_1$ and $p_2$ are incoming momenta and ${p_1'}$ and ${p_2'}$ outgoing momenta with $p_1^2 \equiv {p_1'}^2 \equiv m_1^2$ and $p_2^2\equiv {p_2'}^2\equiv m_2^2$. We employ standard Mandelstam conventions throughout this presentation:
\begin{equation}\label{e:sdef} s\equiv(p_1+p_2)^2 \equiv({p_1'}+{p_2'})^2 
= m_1^2+m_2^2+2m_1 m_2 \sigma,\quad \sigma \equiv \frac{p_1 \cdot p_2}{m_1 m_2}\,, \end{equation} 
\begin{equation}\label{e:tdef}
t \equiv (p_1-{p_1'})^2\equiv ({p_2'}-p_2)^2\equiv q^2=-\vec{q}^{\,2}\,,\end{equation}
and 
\begin{equation}\label{e:udef}
u \equiv (p_1-{p_2'})^2 \equiv ({p_1'}-p_2)^2\,,
\end{equation}
where $s$ defines the center of mass energy, $E_{CM}^2$, and $t$ is  the transfer momentum. One sees from the definitions that 
\begin{equation} p_1\cdot q=\frac{q^2}{2}\,,\ p_2\cdot q=-\frac{q^2}{2}\,,\ {p_1'}\cdot q=-\frac{q^2}{2}\,, \ {p_2'}\cdot q=\frac{q^2}{2}\,.\end{equation}
%
\subsection{Second Post-Minkowskian order in Einstein gravity}\label{sec:2PMlevel}
In this section we recompute the classical contributions to the
one-loop amplitude in Einstein gravity.  This has
already been done in many previous works but the method used here will
be a good propaedeutic for the two-loop analysis that will be done in
later in the paper.\\[5pt]
We can expand the one-loop amplitude in Einstein gravity in terms of basis
integrals
\begin{equation}\label{e:M1loop}
\mathcal{M}^{\rm 1-loop}(p_1,p_2,p_1',p_2')=i64 \pi^2
    G_N^2\left(c_{\Box}\mathcal{I}_{\Box}+ c_{\bowtie} \mathcal{I}_{ \bowtie}+
c_{\triangleright}\mathcal{I}_{\triangleright}+c_{\triangleleft}\mathcal{I}_{\triangleleft} + \ldots\right),
\end{equation}
spanned  by the scalar box $I_{\Box}$, the scalar cross-box
$I_{\bowtie} $, and scalar triangles $I_{\triangleleft}$ integral
functions. The remainder in the amplitude are bubble integral 
and rational
functions which  do not contribute to the classical limit and give quantum mechanical contributions that we neglect.
At leading order in the momentum transfer $\vec q$ the coefficients in
$D$ dimensions are given by~\cite{KoemansCollado:2019ggb} (after
reducing the quadratic triangle on scalar triangles)
\begin{align}\label{e:coeff1loop}
  c_{\Box}&=c_{\bowtie}= 16m_1^4m_2^4 {(1-(D-2)\sigma^2)^2\over(D-2)^2}, \cr
c_{\triangleright}&=\frac{4 m_1^4 m_2^2 \left(D-7 +(D (4 D-17)+19)\sigma^2 \right)}{(D-2)^2}, \cr
  c_{ \triangleleft}&=\frac{4 m_1^2m_2^4 \left(D-7 +(D (4 D-17)+19)\sigma^2\right)}{(D-2)^2}\,.
\end{align}
We consider the classical limit of the box and triangle integrals.
\subsubsection{Scalar triangle integrals}
We consider the classical limit of the scalar  triangle integral
\begin{equation}
\mathcal{I}_{\triangleright}(q^2)=\int \frac{d^D\ell}{(2
  \pi\hbar)^D}\frac{\hbar^4}{((\ell+q)^2+i\varepsilon)(\ell^2+i\varepsilon)((\ell+p_1)^2-m^2_1+i\varepsilon)}\,,
\end{equation}
we symmetrise this expression over $p_1$ and ${p_1'}$ 
\begin{equation}\begin{split}
\mathcal{I}_{\triangleright}(q^2)&={1\over 2\hbar^{D-4}}\int \frac{d^D\ell}{(2
  \pi)^D}\frac{1}{((\ell+q)^2+i\varepsilon)(\ell^2+i\varepsilon)} \, \big({1\over (\ell+p_1)^2-m^2_1+i\varepsilon}\\
  & \hskip8cm
+{1\over (\ell-{p_1'})^2-m^2_1+i\varepsilon}\big)\,.
\end{split}\end{equation}
Using that $p_1^2={p_1'}^2=m_1^2$ and the notation $q=|\vec q|\, u_q$ so
that $u_q^2=-1$, and rescaling the loop momentum $\ell=|\vec q| l$,
and   $\vec q=\hbar \underline{\vec q}$, we have at leading order in $q^2$
\begin{equation}
\mathcal{I}_{\triangleright}(q^2)={|\underline q|^{D-5}\over 2\hbar}\int \frac{d^Dl}{(2
  \pi)^D}\frac{1}{((l+u_q)^2+i\varepsilon)(l^2+i\varepsilon)}
\left({1\over 2l\cdot p_1+i\varepsilon}-{1\over 2l\cdot p_1-i\varepsilon}\right)\,.
\end{equation}
In this expression we have approximated the propagator
\begin{equation}
  {1\over (\ell+p_1)^2-m_1^2+i\varepsilon}= {1\over \ell^2+2\ell\cdot
    p_1+i\varepsilon}={1\over \hbar^2 l^2 |\underline q|^2
    +2\hbar |\underline q| l\cdot
    p_1+i\varepsilon}\simeq {1\over 2\hbar|\underline q| l\cdot
    p_1+i\varepsilon},
\end{equation}
since the classical contribution will come from the leading order in
$\hbar$ of this expression.\\[5pt]
Using the definition of the delta-function as a
distribution\footnote{Our 
  prescription for the propagators is ${i\over p^2-m^2+i\varepsilon}=
  PP\left(i\over p^2-m^2\right)-i\pi \delta(p^2-m^2)$
  with $\varepsilon>0$ and the mostly minus signature
  $(+-\cdots-)$. This, together with the appropriate contour of
  integrations (cf.~page 31~of~\cite{Peskin:1995ev}), is
the way to regulate the pole in the Green function. }
\begin{equation}\label{e:PP}
 \lim_{\varepsilon\to0^+}\left( {1\over x-i\varepsilon}-{1\over x+i\varepsilon}\right)=
 \lim_{\varepsilon\to0^+} {2i\varepsilon \over x^2+\varepsilon^2}=2i\pi 
\delta(x).
\end{equation}
We then have in the limit $\varepsilon \to0$ that
\begin{equation}
\mathcal{I}_{\triangleright}(q^2)=- {|\underline q|^{D-5}\over 2\hbar}\int \frac{d^Dl}{(2
  \pi)^D}\frac{2i\pi \delta(2l\cdot p_1)}{((l+u_q)^2+i\varepsilon)(l^2+i\varepsilon)}+{\cal O}(\varepsilon)\,.
\end{equation}
The integral reduces to the scalar bubble integral in $D-1$ dimensions
\begin{equation}
  \mathcal I_{\triangleright}(q^2)=-\frac{i |\underline q|^{D-5}}{4 m_1\hbar}\int \frac{d^{D-1}
  l}{(2\pi)^{D-1}}\frac{1}{(l^2+i\varepsilon) ((l+u_q)^2+i\varepsilon)}+{\cal O}(\varepsilon)\,,
\end{equation}
which can be easily computed to give
\begin{equation}
\mathcal I_{\triangleright}(q^2)=\frac{|\underline q|^{D-5}}{4 m_1\hbar} \frac{\Gamma(\frac{D-3}{2})\Gamma(\frac{5-D}{2})}{(4\pi)^{\frac{D-1}{2}}\Gamma(D-3)} \,.
\end{equation}
In $D=4-2\epsilon$ we get 
\begin{equation}
I_{\rhd}(q^2)=\frac{1}{32 m_1 \hbar |\underline q|}+ {\cal O}(\epsilon)\,.
\end{equation}

\subsubsection{Box integral}
The scalar box integral is defined as
\begin{equation}
\mathcal I_{\Box}=\int \frac{d^D \ell}{(2\pi \hbar)^D} \frac{\hbar^4}{((\ell+p_1)^2-m_1^2+i \varepsilon)((\ell-p_2)^2-m_2^2+i \varepsilon)(\ell^2+i\varepsilon) ((\ell+q)^2+i\varepsilon)}\,,
\end{equation}
and the scalar cross-box is 
\begin{equation}
\mathcal I_{\bowtie}=\int \frac{d^D \ell}{(2\pi \hbar)^D} \frac{\hbar^4}{((\ell+p_1)^2-m_1^2+i \varepsilon)((\ell+p_2')^2-m_2^2+i \varepsilon)(\ell^2+i\varepsilon) ((\ell+q)^2+i\varepsilon)}\,.
\end{equation}
After symmetrization on the external legs the sum of the
box and the cross-box contributions
$\mathcal I_{\boxtimes}=\mathcal I_{\Box}+\mathcal I_{\bowtie}$ reads
\begin{multline}
\mathcal I_{\boxtimes}=\int \frac{d^D \ell}{(2\pi \hbar)^D}
\frac{\hbar^4}{(\ell^2+i\varepsilon)
  ((\ell+q)^2+i\varepsilon)}\left({1\over (\ell+p_1)^2-m_1^2+i \varepsilon} +{1\over
    (\ell-p_1')^2-m_1^2+i \varepsilon} \right) \cr
\times\left( {1\over (\ell-p_2)^2-m_2^2+i \varepsilon}+{1\over (\ell+p_2')^2-m_2^2+i \varepsilon}\right)\,.
\end{multline}
As before we change variables $\ell \rightarrow |\vec{q}| l$ with  
$\vec q=\hbar \underline{\vec q}$  to get
\begin{multline}
\mathcal I_{\boxtimes}={|\underline{\vec q}|^{D-6}\over \hbar^2}\int \frac{d^D l}{(2\pi)^D}
\frac{1}{(l^2+i\varepsilon)
  ((l+ u_q)^2+i\varepsilon)}\cr
\times\left({1\over 2l\cdot p_1+\hbar l^2 |\underline{\vec q}|+i \varepsilon} +{1\over
    -2l\cdot p_1+\hbar |\underline{\vec q}| (l+u_q)^2 +i \varepsilon} \right) \cr
\times\left( {1\over -2l\cdot p_2+\hbar l^2 |\underline{\vec q}|+i \varepsilon}+{1\over
    2l\cdot p_2+\hbar |\underline{\vec q}| (l+u_q)^2 +i \varepsilon}\right)\,,
\end{multline}
where we used that  ${p_1'}=p_1-q$ and ${p_2'}=p_2+q$.
As the box contribution is of order $1/\hbar^2$ one needs to do the
small $q=\hbar\underline q$ expansion of the integral of the second
order 
\begin{equation}\label{e:boxexpansion}
 \mathcal I_{\boxtimes}={|\underline{\vec q}|^{D-6}\over \hbar^2}\Big(\mathcal I_{\boxtimes}^0+
 \hbar |\underline{\vec{q} }| \mathcal I_{\boxtimes}^1+(\hbar
 |\underline{\vec{q} }|)^2 \mathcal I_{\boxtimes}^2+ O((\hbar
 \underline q)^3)\,\Big).
\end{equation}
For doing this expansion we change variables to $p_1=\bar
p_1+{\hbar \underline q\over2}$ and $p_2=\bar p_2-{\hbar \underline q\over2}$.

\paragraph{The leading contribution}
\label{sec:leading-term}
is given by 
\begin{align}
\mathcal I_{\boxtimes}^0=-\frac{1}{2}\int
\frac{d^D l}{(2\pi)^{D-2}} 
\frac{\delta(2\bar p_1\cdot l) \delta(2\bar p_2\cdot l)}{(l^2+i\varepsilon) ((l+u_q)^2+i\varepsilon)}+{\cal O}(\varepsilon)\,,
\end{align}
where we made use of the result in eq.~\eqref{e:PP}.
The delta-function constraints imply to divide by the determinant of the two vectors $
{(\epsilon_{\mu_1 \mu_2} \bar p_1^{\mu_1} \bar p_2^{\mu_2})^2}={(\bar
  p_1\cdot \bar p_2)^2-{\bar p_1^2} {\bar p_2^2}}=m_1^2m_2^2\left(\sigma^2-1- {(\hbar
  |{\underline q}|)^2s  \over 4m_1^2m_2^2}\right)$, to give
\begin{equation}\displaystyle
\mathcal I_{\boxtimes}^0=-\frac{1}{8
  m_1m_2\sqrt{\sigma^2-1-  {(\hbar
  |{\underline q}|)^2s  \over 4m_1^2m_2^2}}}\int \frac{d^{D-2} l}{(2\pi)^{D-2}} \frac{1}{(l^2+i\varepsilon) ((l+u_q)^2+i\varepsilon)}+{\cal O}(\varepsilon)\,,
\end{equation}
using the result given in Appendix~A of~\cite{SmirnovEvaluating}
\begin{equation}
\int \frac{d^{D-2} l}{l^2 (l+u_q)^2}=\frac{ \pi^{\frac{D-2}{2}}\Gamma(\frac{D-4}{2})^2\Gamma(\frac{6-D}{2})}{\Gamma(D-4)}\,,
\end{equation}
we have for the leading order
\begin{equation}\label{e:Ibox0}
\mathcal I_{\boxtimes}^0=-\frac{1}{8 m_1m_2 \sqrt{\sigma^2-1 - {(\hbar
  |{\underline q}|)^2s  \over 4m_1^2m_2^2}}}\frac{ \pi^{\frac{D-2}{2}}\Gamma(\frac{D-4}{2})^2\Gamma(\frac{6-D}{2})}{(2\pi)^{D-2}\Gamma(D-4)}\,.
\end{equation}
%
\paragraph{The next-to-leading contribution}
\label{sec:subleading-term}
in the expansion, is obtained after 
applying eq.~\eqref{e:PP}, and reads
\begin{multline}\label{e:Ionedelta}
  \mathcal I_{\boxtimes}^1=\frac{i}{2}\int \frac{d^D l}{(2\pi)^{D-1}}
\frac{1}{(l^2+i\varepsilon)
 ((l+u_q)^2+i\varepsilon)}\cr
\times  \Big(\delta(2\bar p_2\cdot l)\left(\frac{u_q.l}{(2\bar p_1 \cdot
  l-i\varepsilon)^2}+\frac{u_q.l}{(2\bar p_1 \cdot
  l+i\varepsilon)^2}\right)\cr
+\delta(2\bar p_1\cdot l)\left(\frac{u_q.l}{(2\bar p_2 \cdot
  l+i\varepsilon)^2}+\frac{u_q.l}{(2\bar p_2 \cdot
  l-i\varepsilon)^2}\right)\Big)+{\cal O}(\varepsilon)\,,
\end{multline}
neglecting the  tadpole contribution which do not contribute to the classical limit 
we get
\begin{equation}
\!\!  \mathcal I_{\boxtimes}^1\!\simeq\! \frac{i}{2}\!\!\int \!\!\frac{d^D l}{(2\pi)^{D-1}}
{1\over  (l^2+i\varepsilon) ((l+u_q)^2+i\varepsilon)}\left(\!\frac{\delta(2\bar p_2\cdot l)}{(2\bar p_1 \cdot
  l-i\varepsilon)^2}+\frac{\delta(2\bar p_1\cdot l)}{(2\bar p_2 \cdot
  l+i\varepsilon)^2}\!\right)+{\cal O}(\varepsilon)\,.
\end{equation}
In the first integral we choose a vector $v$, with
$v^2=-1$, orthogonal to $\bar p_2$ in
the plane generated by $\bar p_1$ and $\bar p_2$.  Then we have $\bar p_1= 
{m_1\sigma \bar p_2\over m_2}-m_1
\sqrt{\sigma^2-1}\,v$, such that $\bar p_1\cdot l= {m_1\sigma \bar p_2\cdot l\over m_2}-m_1
\sqrt{\sigma^2-1}\,v\cdot l$. Therefore in the first integral we have
\begin{equation}
    \frac{\delta(2\bar p_2\cdot l)}{(2\bar p_1 \cdot
  l-i\varepsilon)^2}= \frac{\delta(2\bar p_2\cdot
  l)}{(-2m_1\sqrt{\sigma^2-1} v\cdot  l-i\varepsilon)^2}\,.
\end{equation}
Similarly,  in the second integral  we choose a vector $v$, with
$v^2=-1$, orthogonal to $\bar p_1$ in
the plane generated by $\bar p_1$ and $\bar p_2$. Then we have $\bar p_2= 
{m_2\sigma \bar p_1\over m_1}-m_2
\sqrt{\sigma^2-1}\,v$  and $\bar p_2\cdot l= {m_2\sigma \bar p_1\cdot l\over m_1}-m_2
\sqrt{\sigma^2-1}\,v\cdot l$ so that
\begin{equation}
  \frac{\delta(2\bar p_1\cdot l)}{(2\bar p_2 \cdot
  l+i\varepsilon)^2}=
 \frac{\delta(2\bar p_1\cdot l)}{(-2m_2 \sqrt{\sigma^2-1} v\cdot
  l+i\varepsilon)^2}\,.
\end{equation}
 Taking into account the delta-function, we
have 
\begin{equation}
\!\!\mathcal I_{\boxtimes}^1\!=\!\frac{i(m_1+m_2)
}{16 m_1^2m_2^2\left(\sigma^2-1\right)}\int\!\! \frac{d^{D\!-\!1}
  l}{(2\pi)^{D\!-\!1}} \frac{1}{(v\cdot l)^2 (l^2+i\varepsilon)
  ((l+u_q)^2+i\varepsilon)}+{\cal O}(\varepsilon, \underline q^2)\,.
\end{equation}
Using the result from Appendix~A of~\cite{SmirnovEvaluating}
\begin{equation}
\int \frac{d^{D-1} l}{(p\cdot l)^2 l^2 (l+u_q)^2}
=-\frac{2 \pi^{\frac{D-1}{2}}\Gamma(\frac{D-5}{2})^2 \Gamma(\frac{7-D}{2})}{\Gamma(D-5)}=\frac{4 \pi^{\frac{D-1}{2}}\Gamma(\frac{D-3}{2})^2 \Gamma(\frac{5-D}{2})}{\Gamma(D-4)}\,,
\end{equation}
we get that
\begin{equation}\label{e:Ibox1}
\mathcal I_{\boxtimes}^1=\frac{i(m_1+m_2)}{16 m_1^2m_2^2\left(\sigma^2-1-  {(\hbar
  |{\underline q}|)^2s  \over 4m_1^2m_2^2}\right)}\frac{4 \pi^{\frac{D-1}{2}}\Gamma(\frac{D-3}{2})^2 \Gamma(\frac{5-D}{2})}{(2\pi)^{D-1}\Gamma(D-4)}\,.
\end{equation}
%

\paragraph{The next-to-next-leading contribution}
\label{sec:sub-sub-leading}
is  of order $(\hbar |{\underline q}|)^2$. This is obtained by expanding the massive propagators
\begin{eqnarray} 
\mathcal I_{\boxtimes}^2&&=\!\!\!\!\!\!\!\!\!\!\!\!\!\!\!\!\!\!\!\!\!\\ 
&&\!\!\!\!\!\!\!\!\!\!\!\frac{-1}{8 \hbar^2}\!\!\int\!\!\! \frac{d^D
  l}{\left(2\pi\right)^D}\!\! \left(\frac{\frac14\left(u_q \cdot
      l\right)^2}{\left(\bar{p}_1 \cdot
      l\!-\!i\varepsilon\right)^3}-\frac{\frac14\left(u_q \cdot
      l\right)^2}{\left(\bar{p}_1 \cdot
      l\!+\!i\varepsilon\right)^3}\right)\!\!\left(\frac{1}{\bar{p}_2 \cdot
    l\!+\!i\varepsilon}-\frac{1}{\bar{p}_2 \cdot
    l\!-\!i\varepsilon}\right)\frac{1}{l^2 \left(l+u_q\right)^2}\cr
&& \!\!\!\!\!\!\!\!\!\!\!\frac{-1}{8 \hbar^2}\!\!\int\!\!\! \frac{d^D l}{\left(2\pi\right)^D}
\!\!\left(\frac{1}{\bar{p}_1 \cdot l\!-\!i\varepsilon}-\frac{1}{\bar{p}_1 \cdot
    l\!+\!i\varepsilon}\right)\!\!\left(\frac{\frac14\left(u_q \cdot
      l\right)^2}{\left(\bar{p}_2 \cdot
      l\!+\!i\varepsilon\right)^3}-\frac{\frac14\left(u_q \cdot
      l\right)^2}{\left(\bar{p}_2 \cdot
      l\!-\!i\varepsilon\right)^3}\right)\frac{1}{l^2
  \left(l\!+\!u_q\right)^2}\cr
&&\!\!\!\!\!\!\!\!\!\!\!\frac{-1}{8 \hbar^2}\!\!\int\!\!\! \frac{d^D l}{\left(2\pi\right)^D} \!\!\left(\frac{-\frac12u_q \cdot l}{\left(\bar{p}_1 \cdot l\!-\!i\varepsilon\right)^2}-\frac{\frac12u_q \cdot l}{\left(\bar{p}_1 
\cdot l\!+\!i\varepsilon\right)^2}\right)\left(\frac{\frac12u_q \cdot l}{\left(\bar{p}_2 \cdot l\!+\!i\varepsilon\right)^2}\!+\frac{\frac12u_q \cdot l}{\left(\bar{p}_2 \cdot l\!-\!i\varepsilon\right)^2}\right)\!\!\frac{1}{l^2 \left(l\!+\!u_q\right)^2}\nonumber\,.
\end{eqnarray}
Using the principal part identity eq.~\eqref{e:PP} and  {\tt
  LiteRed}~\cite{Lee:2013mka} for reducing the integrals to one-loop
master integrals, we obtain
\begin{equation}
\mathcal I_{\boxtimes}^2=   \frac{(4\pi)^{\epsilon}(1+2\epsilon)\Gamma(1+\epsilon)\Gamma(-\epsilon)^2}{64\hbar^2 \pi^2 m_1^2 m_2^2 (\sigma^2-1)^{\frac{3}{2}}\Gamma(-2\epsilon)} \left(\frac{s \pi}{4 m_1 m_2 }+i(\sigma \arccosh(\sigma)-\sqrt{\sigma^2-1})\right)\,.
\end{equation}
%

\paragraph{Summary of the one-loop computation in Einstein gravity}
\label{sec:summary-one-loop}
Summing these contributions one gets  for the scalar box integral contributions in $D=4-2\epsilon$
\begin{multline}
\mathcal
I_{\boxtimes}={1\over|\underline q|^{2+2\epsilon}}\Bigg(-\frac{(4\pi)^{\epsilon}\Gamma(1+\epsilon)\Gamma(-\epsilon)^2}{32
  \hbar^2 \pi m_1 m_2 \Gamma(-2\epsilon)\displaystyle
  \sqrt{\sigma^2-1-{(\hbar
  |{\underline q}|)^2s  \over 4m_1^2m_2^2}}}\cr\hskip6cm\displaystyle
+\frac{i(4\pi)^\epsilon (m_1+m_2)|\vec{q}|}{32 \hbar m_1^2 m_2^2
 \left (\sigma^2-1\right)}\frac{\Gamma(\frac{1}{2}-\epsilon)^2
  \Gamma(\frac{1}{2}+\epsilon)}{\pi^{3\over2}\Gamma(-2\epsilon)}\displaystyle\cr +\displaystyle
\frac{(4\pi)^{\epsilon}|\vec{q}|^2\Gamma(1+\epsilon)\Gamma(-\epsilon)^2}{64 \pi^2 m_1^2 m_2^2 \displaystyle(\sigma^2-1)^{\frac{3}{2}}\Gamma(-2\epsilon)} \left(\!\frac{s \pi(1\!+\!2\epsilon)}{4 m_1 m_2 }\!+\!i(1\!+\!2\epsilon)(\sigma \arccosh(\sigma)\!-\!\sqrt{\sigma^2\!-\!1})\!\right)\!+\!{\cal O}(|\vec{q}|^3)\Bigg)\,.
\end{multline}
Expanding the denominator of the first term leads to\footnote{This
  expansion is justified as long as $\sigma\neq 1$, which is the
  hypothesis we will use since we are considering the scattering
  angle. In the static case $\sigma=1$  the amplitude reproduces the
  one-loop computation of~\cite{Bjerrum-Bohr:2013bxa}.} 
\begin{multline}\label{e:boxexpand}
\mathcal
I_{\boxtimes}={1\over|\underline q|^{2+2\epsilon}}\Bigg(-\frac{(4\pi)^{\epsilon}\Gamma(1+\epsilon)\Gamma(-\epsilon)^2}{32
  \hbar^2 \pi m_1 m_2 \Gamma(-2\epsilon)
  \sqrt{\sigma^2-1}}+\frac{i(4\pi)^\epsilon (m_1+m_2)|\vec{q}|}{32
  \hbar m_1^2 m_2^2 (\sigma^2-1)}\frac{\Gamma(\frac{1}{2}-\epsilon)^2
  \Gamma(\frac{1}{2}+\epsilon)}{\pi^{3\over 2}\Gamma(-2\epsilon)}+\cr
\frac{(4\pi)^{\epsilon}|\vec{q}|^2\Gamma(1+\epsilon)\Gamma(-\epsilon)^2}{64 \pi^2 m_1^2 m_2^2 (\sigma^2-1)^{\frac{3}{2}}\Gamma(-2\epsilon)} \left(\frac{\epsilon s \pi }{2 m_1 m_2 }\!+\!i(1\!+\!2\epsilon)(\sigma \arccosh(\sigma)\!-\!\sqrt{\sigma^2-1})\right)+{\cal O}(|\vec{q}|^3)\Bigg)\,.
\end{multline}
This expression matches the sum of the  expressions given  in
eq.~(B.36) and~(B.40) in~\cite{Cristofoli:2020uzm} and  eq.~(4.54)
and~(4.59) in~\cite{Parra-Martinez:2020dzs} up to a normalization constant.
The result is valid in arbitrary dimension, and  it
keeps covariant vectors everywhere.
\subsubsection{Comparing the soft and the potential region}
\label{sec:comp-soft-potent}
%
We can now compare the result for the box to the one derived from the
potential region and see why this is leads to a different
answer. 
In order to do the potential region versus the  soft region analysis,
we will define
\begin{equation}\label{e:OmegaDef}
  \Omega=\frac{p_1+p_2}{\sqrt{s}}\,,
\end{equation}
which is a time-like vector orthogonal to $q$. We will also define the
vector
\begin{equation}\label{e:vdef}
  \tilde{v}_\mu=\frac{\epsilon_{\mu\nu\rho\sigma} p_1^\nu p_2^\rho u_q^\sigma}{m_1 m_2},
\end{equation}
 orthogonal to $p_1$, $p_2$, and $q$, and $v$
 a unit vector ($v^2=-1$) orthogonal to $\Omega$, $u_q$, and
 $\tilde{v}$.
 The vectors $(\Omega, u_q, v, \tilde{v})$ form a basis of the four
 dimensional Minkowski space, on which one can expand $p_1$ and $p_2$ as
\begin{equation}
p_1=\frac{m_1^2+m_1 m_2 \sigma}{\sqrt{s}} \Omega+\frac{|\vec{q}|}{2} u_q +  v\sqrt{\frac{m_1^2 m_2^2(\sigma^2-1)}{s}-\frac{|\vec{q}|^2}{4}}\,,
\end{equation}
and
\begin{equation}
p_2=\frac{m_2^2+m_1 m_2 \sigma}{\sqrt{s}} \Omega-\frac{|\vec{q}|}{2} u_q - v \sqrt{\frac{m_1^2 m_2^2(\sigma^2-1)}{s}-\frac{|\vec{q}|^2}{4}}\,.
\end{equation}
The potential region is defined as the integration region where the
time component of the loop momentum, $\omega\equiv \Omega\cdot l$ is very 
small compared to the spatial components.
This will not have any consequences in the evaluation of the
leading contribution $\mathcal I_\boxtimes^0$ in eq.~\eqref{e:Ibox0}  because the delta-functions in this integral always cancel the time component in the graviton propagators.
But this will have effects on the sub-leading contribution $\mathcal
I_\boxtimes^1$ in~\eqref{e:Ionedelta}. In the
potential region this contribution can be written  at leading
order in $|\underline q|$ as
\begin{eqnarray}
\mathcal I_{\boxtimes}^{1p}=-\frac{i |\underline q|^{D-6}}{4
  \hbar}\int \frac{d^D l}{(2\pi)^{D}} {1\over
  (\vec{l}^2-i\varepsilon) ((\vec{l}+\vec{u_q})^2-i\varepsilon)} && \\\ &&\hskip-9cm\displaystyle
\times\left(\displaystyle\frac{\displaystyle\delta\Big(\frac{m_2^2+m_1 m_2 \sigma}{\sqrt{s}}\omega-\sqrt{\frac{m_1^2 m_2^2(\sigma^2-1)}{s}}v\cdot l\Big)}{\Big(\displaystyle 2\frac{m_1^2+m_1 m_2 \sigma}{\sqrt{s}}\omega+2\sqrt{\frac{m_1^2 m_2^2(\sigma^2-1)}{s}}v\cdot l-i\varepsilon\Big)^2}\right. \cr && \hskip-4cm+\left.\displaystyle \frac{\displaystyle\delta\Big(\frac{\displaystyle m_1^2+m_1 m_2 \sigma}{\displaystyle\sqrt{\displaystyle 
s}}\omega+\sqrt{\frac{m_1^2 m_2^2(\sigma^2-1)}{s}}v\cdot l\Big)}{\Big(2\frac{\displaystyle m_2^2+m_1 m_2 \sigma}{\displaystyle \sqrt{\displaystyle 
s}}\omega-2\sqrt{\displaystyle\frac{m_1^2 m_2^2(\sigma^2-1)}{s}}v\cdot l+i\varepsilon\Big)^2}\right)\nonumber\,.
\end{eqnarray}
Integrating over the $\omega$ variable gives 
\begin{eqnarray}
\mathcal I_{\Box}^{1p}=-\frac{i |\underline q|^{D-6}}{4
  \hbar}&&\int \frac{d^{D-1} l}{(2\pi)^{D-1}}
{1\over  (\vec{l}^2-i\varepsilon)
  ((\vec{l}+\vec{u_q})^2-i\varepsilon)} \\ &&\hskip-2cm \nonumber
\times\left(\displaystyle{{\displaystyle \sqrt s\over \displaystyle m_2^2+m_1m_2\sigma}\over \left(2 {\displaystyle\sqrt{\displaystyle s  m_1^2 m_2^2(\sigma^2-1)}\over \displaystyle m_2^2+m_1m_2\sigma}v\cdot l-i\varepsilon\right)^2}  +{{\displaystyle\sqrt s\over \displaystyle m_2^2+m_1m_2\sigma}\over \left(2 {\displaystyle\sqrt{s  m_1^2 m_2^2(\sigma^2-1)}\over \displaystyle m_1^2+m_1m_2\sigma}v\cdot l+i\varepsilon\right)^2}\right)\,,
\end{eqnarray}
leading to 
\begin{equation}
\mathcal I_{\Box}^{1p}=\frac{i  |\underline q|^{D-6}\sqrt{s}}{16 \hbar m_1^2
  m_2^2(\sigma^2-1)}\int \frac{d^{D-1} l}{(2\pi)^{D-1}}
\frac{1}{(v\cdot l)^2  (\vec{l}^2-i\varepsilon)
  ((\vec{l}+\vec{u_q})^2-i\varepsilon)}\,.
\end{equation}
Since $\displaystyle\sigma={p_1\cdot p_2\over m_1m_2}$ we find that this integral
is related to the box contribution in eq.~\eqref{e:Ibox1} as
\begin{equation}
{\mathcal I_{\Box}^{1p}=\frac{\sqrt{s}}{m_1+m_2}\mathcal I_{\Box}^{1}}\,.
\end{equation}
This is exactly the difference noticed
in eq.~(B.57)~\cite{Cristofoli:2020uzm} between the evaluation in the
potential region and the
soft region at one-loop.\\[5pt]
Actually the difference  between the potential region $\mathcal I_{\Box}^{1p}$ is
and the full (soft) integral $\mathcal I_{\boxtimes}^{1}$ is the
expansion of the internal graviton propagator with respect to the $\omega$
\begin{multline}
\mathcal I_{\Box}^{1}-\mathcal I_{\Box}^{1p}=\frac{i}{4 \hbar}\int
\frac{d^D l}{(2\pi)^{D-1}} \left(\frac{\delta(p_2\cdot l)}{(2p_1 \cdot
  l-i\varepsilon)^2}+\frac{\delta(p_1 \cdot l)}{(2p_2\cdot
  l+i\varepsilon)^2}\right)\cr
\times\sum_{n=1}^{\infty}
\sum_{l=0}^{n}\frac{\omega^{2n}}{(\vec{l}^2-i\varepsilon)^{(l+1)}
  ((\vec{l}+\vec{u_q})^2-i\varepsilon)^{(n-l+1)}} \,.
\end{multline}
Although the exchange of the velocity expansion and integration does
not lead to a different result at one-loop order,  this will be
different at the two-loop order. This will be apparent when comparing
 the potential region result with the amplitude computation
done in this work.
\section{Two-body amplitudes in maximal supergravity}\label{sec:3PMlevel}
In this section, we summarize the evaluation of four-point maximal supergravity amplitudes with
massive external states up to third order in Newton's constant.\\[5pt]
The generic maximal supergravity four-point amplitude
takes the form of a helicity dependent kinematical factor $\mathbf
 R^4_{\zeta_1,\zeta_2,\zeta_3,\zeta_4}$ times a sum of loop integrals
 that does
 not depend on the helicities $\zeta_i$ of the external states 
\begin{equation}\begin{split}\label{e:MN8max}
\hskip-9cm \mathcal M^{N=8}(\zeta_1,\dots,\zeta_4)=  \mathbf
 R^4_{\zeta_1,\zeta_2,\zeta_3,\zeta_4}\times &\\ & \hskip-4cm \Big[\mathcal 
 M_0 (p_1,\dots,p_4)\!+\!\mathcal
  M_1 (p_1,\dots,p_4)\!+\!\ldots \! +\! \mathcal M_L (p_1,\dots,p_4)\! +\! {\cal O}(G_N^{L+2})\Big]\,.\!\!\!\!\!\!\!\!\!\!\!\!\!\!\!\!\!\!\!\!\!\!\!\!\!\!\!\!\!\!\! \hskip-3cm
\end{split}\end{equation}
Following eq.~(7.4.57) of~\cite{GSW1} we define
\begin{equation}
 \mathbf R^4_{\zeta_1,\zeta_2,\zeta_3,\zeta_4}\equiv\zeta_1^{AA'}
 \zeta_2^{BB'} \zeta_3^{CC'} \zeta_4^{DD'} K_{ABCD} \tilde K_{A'B'C'D'}\,,
\end{equation}
where the indices $A$, $B$ on the polarization tensor $\zeta^{AB}_r$
run over vector and spinor values. The tensors $K$ and $\tilde K$ 
are expressed as traces of the fermionic zero
modes as shown in Appendix~9.A of~\cite{GSW2}.
For external graviton states the tensors $K$ and
$\tilde K$ are given by the expression in eq.~(9.A.18) of~\cite{GSW2} and 
one obtains the
linearization of the $R^4$-kinematic factor and $\mathbf
R^4_{\zeta_1,\zeta_2,\zeta_3,\zeta_4}=\prod_{r=1}^4
\hat R^{(r)\mu_{2r-1}\mu_{2r}\nu_{2r-1}\nu_{2r}}t^8_{\mu_1\cdots
 \mu_8}t^8_{\nu_1\cdots \nu_8}$ where $\hat
R_{\mu\nu\rho\sigma}=-4p_{[\mu}\zeta_{\nu][\rho} p_{\sigma]}$ is the
linearized Riemann tensor.
We obtain external massive states by considering Kaluza-Klein
reduction~\cite{Caron-Huot:2018ape}. The kinematic variable then
becomes
\begin{equation}\label{e:KKreduc}
 S=s-m_1^2-m_2^2=2m_1m_2\sigma, \, T=t=\hbar^2 \underline q^2,
 \, U=u-m_1^2-m_2^2=-2m_1m_2\sigma-\hbar^2\underline q^2 \,.
\end{equation}
Since this construction preserves
maximal supersymmetry, the helicity dependence in eq.~\eqref{e:MN8max} is
preserved. 
For the choice of helicity states
in~\cite{DiVecchia:2021ndb,DiVecchia:2020ymx} the kinematical factor
is
\begin{multline}
 \label{e:KV}
 \mathbf R^4_{\phi,\phi,\varphi,\varphi}=\frac12 (S^4+T^4+U^4) =16 m_{1}^4 m_{2}^4 \sigma
 ^4-16 m_{1}^3 m_{2}^3 \underline q^2
 \sigma ^3\hbar^2+
 12 m_{1}^2 m_{2}^2 (\underline q^2)^2 \sigma ^2\hbar^4\cr
 -4 m_{1} m_{2} \sigma (\underline q^2)^3\hbar^6
  + (\underline q^2)^4\hbar^8\,,
\end{multline}
whereas for the choice of helicities
in~\cite{Caron-Huot:2018ape,Parra-Martinez:2020dzs} the kinematical
factor is given by
\begin{equation}
 \label{e:KP}
 \mathbf R^4_{\phi,\phi,\phi,\phi}=S^4=16 m_1^4m_2^4\sigma^4\,,
\end{equation}
where we used the kinematical relations
in eq.~\eqref{e:sdef}--\eqref{e:udef} and the definition of the momentum
transfer $\vec q\equiv\hbar\vec{\underline q}$  tailored for extracting classical physics from amplitudes.\\[5pt]
Focussing on small
momentum transfer $|\underline q| \ll m_1,m_2$ the $L$-order loop amplitude takes the form
\begin{equation}\label{e:genericExpansion}
 \mathcal M_L(\sigma,|\underline q|)= {1\over \hbar^{L-1}|{\underline
   q}|^{2L\epsilon}} \sum_{r\geq-2} \mathcal M_L^{(r)}(\sigma,\epsilon)
 (\hbar |\underline q|)^r .
\end{equation}
We justify the expansion by the following counting. The leading
singular term in the classical limit $\hbar\to0$ arises from the
ladder and cross-ladder diagrams at $L$-loop. These diagram have the
maximal number of massive propagators, with $2L$
massive propagators and $L+1$ massless gravitons exchanges, so the
loop integrals symbolically read (see section~\ref{sec:2PMlevel}
and~\ref{sec:3PMlevel} for details about this at one-loop and
two-loop order amplitude respectively)
\begin{equation}
  \mathcal M_L(\sigma,|q|)|_{\rm ladder}\sim \int \prod_{i=1}^L
  {d^D\ell\over(2\pi \hbar)^D} \, {\hbar^{3L+1}\over
   (2\ell_{i}\cdot p_{j})^{2L}
   (\ell_i^2)^{L+1} } ,
 \end{equation}
 and the classical limit is obtained for $\hbar \to0$ keeping
 $q/\hbar=\underline q$ constant. Rescaling the loop momenta by
 $\ell= \hbar \underline q\, l$ we get
\begin{equation}
  \mathcal M_L(\sigma,|q|)|_{\rm ladder}\sim {1\over \hbar^{L-1}
   |\underline q|^{2L\epsilon} }{1\over (\hbar\underline q)^2} \int {\prod_{i=1}^L
  d^Dl\over
   (2l_{i}\cdot p_{j})^{2L}
   (l_i^2)^{L+1} } .
 \end{equation}
 At $L$-loop order, the other diagrams have more massless propagators
 and less massive propagators. Replacing a massive propagator by a
 massless propagator amounts to multiplying the $\hbar$ count by 
 a factor of $\hbar \underline q$ giving the expansion
 in eq.~\eqref{e:genericExpansion}.
The least number of massive propagator at $L$-loop order is two, so with $3L-1$ massless propagators we have the  $\hbar$ counting
\begin{equation}
 {1\over \hbar^{L-1} |\underline q|^{2L\epsilon} }\,{1\over (\hbar
 |\underline q|)^{3L-1}}\,.
\end{equation}
We note that such a contribution will come from the $t$-channel with an extra
 power of the momentum in the numerator contributing to $(\hbar
 |q|)^{2L}$. This will be the case for the $J_t$ contribution at two-loop 
in eq.~\eqref{e:M2loop}.

\subsection{The amplitudes in momentum space}\label{sec:2PMlevel}
We give the expression of the tree-level amplitude in~\ref{sec:tree},
one-loop amplitude  in~\ref{sec:one}, and two-loop amplitude in~\ref{sec:twoloopresults}.

\subsubsection{Tree-level scalar amplitude}\label{sec:tree}
The helicity independent factor in the massless maximal supergravity four-point tree-level amplitude is given by
\begin{equation}
\mathcal M_0(p_1,\dots,p_4)=\frac{ 8\pi G_N\hbar}{STU}\,,
\end{equation}
which after Kaluza-Klein reduction in eq.~\eqref{e:KKreduc} gives the two-body scattering amplitude
\begin{equation}\label{e:M0result}
\mathcal M_0(\sigma,|\underline q|)=-\frac{8\pi G_N}{2 m_1 m_2 \hbar|\underline q|^2
 \sigma (\hbar^2|\underline q|^2-2 m_1 m_2 \sigma)}=\hbar
{\mathcal M_0^{(-2)}(\sigma)\over \hbar^2|\underline q|^2}+{\cal O}((\hbar
 |\underline q|)^{0})\,,
 \end{equation}
 where
 \begin{equation}
 \mathcal M_0^{(-2)}(\sigma)={ 2\pi G_N\over m_1^2m_2^2\sigma^2}\,.
 \end{equation}

 \subsubsection{One-loop scalar amplitudes}\label{sec:one}
The helicity independent part of the one-loop amplitude is the sum of
the scalar box and cross-box integrals
\begin{equation}\label{e:MoneN8}
\mathcal M_1(p_1,p_2,p_1',p_2')=-i\, (8\pi G_N)^2 \big( \mathcal{I}_{\Box}+\mathcal{I}_{ \bowtie}\big)\,.
\end{equation}
(This result is dimensional reduction of the maximal supergravity
massless one-loop amplitude in~\cite{Green:1982sw} without the
$(T,U)$-channel contribution which vanishes because of choice of
helicity configuration.)

The boxes have been evaluated in
 section~\ref{sec:2PMlevel} with the result in~\eqref{e:boxexpand}
\begin{equation}\displaystyle
\mathcal M_1(\sigma,|\underline q|)={1\over
 |\underline q|^{2\epsilon}}\bigg({\mathcal
 M^{(-2)}_1(\sigma,\epsilon) \over
 \hbar^2|\underline q|^2}+{\mathcal
 M^{(-1)}_1(\sigma,\epsilon)\over \hbar
 |\underline q|}
+\mathcal M^{(0)}_1(\sigma, \epsilon)+{\cal O}(\hbar|\underline q|)\bigg)\,,\displaystyle
\end{equation}
with the coefficient to all orders in $\epsilon$
\begin{align}\label{e:M1result}\displaystyle
\mathcal M^{(-2)}_1(\sigma,\epsilon)&=\frac{2\pi i G_N^2}{m_1 m_2 \sqrt{\sigma ^2-1}} \frac{ (4\pi)^{ \epsilon} \Gamma (-\epsilon )^2 \Gamma (\epsilon
 +1)}{\Gamma (-2 \epsilon )},\\
 \mathcal M^{(-1)}_1(\sigma,\epsilon)&=\frac{2 \sqrt{\pi} (m_1+m_2)G_N^2}{m_1^2 m_2^2 \left(\sigma ^2-1\right) }\frac{ (4\pi)^{\epsilon} \Gamma
 \left(\frac{1}{2}-\epsilon \right)^2 \Gamma \left(\epsilon
 +\frac{1}{2}\right)}{\Gamma (-2
 \epsilon )}\,,
 \end{align}
 \begin{align}\label{e:0result}
 \mathcal M^{(0)}_1(\sigma,\epsilon)&=\frac{ (4\pi) ^{\epsilon } \Gamma 
(-\epsilon )^2 \Gamma (\epsilon +1)
 }{ \Gamma (-2 \epsilon )}\\
 \nonumber &\times \frac{G_N^2 
 \left(2 m_1 m_2 (\sigma \arccosh(\sigma )-\sqrt{\sigma ^2-1}) (2 \epsilon +1) -i \pi s \epsilon \right)}{2m_1^3
 m_2^3 \left(\sigma ^2-1\right)^{3\over2}}\,.
\end{align}
%

\subsubsection{Two-loop scalar amplitudes}\label{sec:twoloopresults}
The helicity independent part of the two-loop amplitude four
points amplitude is the Kaluza-Klein reduction of the  maximal
supergravity two-loop evaluated in~\cite{Bern:1998ug}, with the
$(T,U)$ sector vanishing as being forbidden by  the helicity
configurations~\cite{Parra-Martinez:2020dzs}
\begin{multline}\label{e:Mtwoloop}
\mathcal M_2(\sigma,|\underline q|)={1\over |\vec{\underline
    q}|^{4\epsilon}\hbar}\Big({\mathcal M^{(-2)}_2(\sigma,\epsilon)\over 
  \hbar^2|\underline q|^2}+{\mathcal
  M^{(-1)}_2(\sigma,\epsilon)\over\hbar|\underline q|} +\mathcal
  M^{(0)}_2(\sigma,\epsilon) +{\cal O}(\hbar |\underline q|)
\Big)\,,
\end{multline}
with\footnote{The expression for $\mathcal M^{(-1)}_2(\sigma,\epsilon)$
  differs by a sign with the one given
  in eq.~(5.8) of~\cite{Parra-Martinez:2020dzs}. The present sign guaranties the
  exponentiation of the two-loop amplitude.}
\begin{align}\label{e:M2resultSC}
\mathcal M^{(-2)}_2(\sigma,\epsilon)&= -\frac{G_N^3\sigma^2
  }{
                           3   ( \sigma^2-1)} {(4\pi)^{1+2\epsilon}\Gamma(-\epsilon)^3\Gamma(1+2\epsilon)\over\Gamma(-3\epsilon)},\\
  \mathcal M^{(-1)}_2(\sigma,\epsilon)&=\frac{4 \sqrt{\pi} i G_N^3(m_1+m_2)\sigma^2}{ m_1
  m_2 (\sigma^2-1)^{\frac{3}{2}}}{(4\pi)^{2\epsilon}\Gamma(\frac12-\epsilon)^2\Gamma(\frac12+2\epsilon)\Gamma(-\epsilon)\Gamma(\frac12-2\epsilon)\over \Gamma(\frac12-3\epsilon)\Gamma(-2\epsilon)},
\end{align}
to all orders in $\epsilon$ and 
\begin{multline}
 \mathcal
  M^{(0)}_2(\sigma,\epsilon) =\frac{2G_N^3(4\pi e^{-\gamma_E})^{2\epsilon}}{\pi m_1 m_2 (\sigma^2-1)^2} \Bigg[ \frac{i\pi(1+2\epsilon)\sigma^2(\sigma \arccosh(\sigma)-\sqrt{\sigma^2-1})}{\epsilon^2}+\frac{\pi^2 s\sigma^2}{2\epsilon m_1 m_2} \\ -\frac{\pi^2 (\sigma^2-1)^{\frac32}\arccosh(\sigma)}{\epsilon}- \frac{i\pi}{\epsilon^2}\bigg(\frac{-1}{4(\sigma^2-1)} \bigg)^{\epsilon} \Bigg( (1+2\epsilon)\sigma^2\sqrt{\sigma^2-1}+\sigma(\sigma^2-2)\arccosh(\sigma)\\+\epsilon\big((\sigma^2-1)^{\frac32}- \sigma(\sigma^2-2)\big) \arccosh^2(\sigma) -\epsilon \sigma(\sigma^2-2) \Li_2\left(2-2\sigma(\sigma+\sqrt{\sigma^2-1})\right) \Bigg) +{\cal O}(1) \Bigg],
\end{multline}
where $(-1)^{\epsilon}=1+i\pi \epsilon+{\cal O}(\epsilon^2)$ and ${\cal 
O}(1)$ is defined so that it is regular function both at $\epsilon=0$ and $\sigma=1$. The details of this evaluation are given in section~\ref{sec:evaluationdoubleboxes}.

\subsection{The amplitudes in $b$-space}
\label{sec:ampl-impact-param}
The amplitude is $b$-space is defined by
\begin{equation}
  \widetilde{\mathcal M_L^{N=8}}(\sigma,b)=\frac{1}{4 E_{\rm c.m.}P}\int_{\mathbb R^{D-2}} \frac{d^{D-2}\vec{\underline q}}{(2\pi)^{D-2}}\mathbf R^4_{\zeta_1,\dots,\zeta_4}\mathcal M^{N=8}_L(p_1,p_2,p_1',p_2') e^{i \vec{\underline q}\cdot\vec{b}}\,,
\end{equation}
where $4 E_{\rm c.m.}P= 4m_1 m_2 \sqrt{\sigma^2-1}$.\\[5pt]
For the choice of helicity
in~\cite{Caron-Huot:2018ape,Parra-Martinez:2020dzs} the kinematic
factor is independent of the momentum transfer $\vec q$ and therefore 
\begin{equation}\label{e:MbHelJ}
\widetilde{\mathcal M^{N=8}_L}(\phi,\phi,\phi,\phi)=\frac{4m_1^3m_2^3\sigma^4}{\sqrt{\sigma^2-1}}\int
\frac{d^{D-2}\vec{\underline q}}{(2\pi)^{D-2}}\mathcal M_L(p_1,p_2,p_1',p_2') e^{i \vec{\underline q}\cdot\vec{b}}\,.
\end{equation}
Using the Fourier transformation
\begin{equation}
  \int {d^{D-2} \vec {\underline q}\over (2\pi)^{D-2}} {e^{i\vec
      {\underline q}\cdot \vec b}\over
  |\vec{\underline q}|^{2\alpha}}= {\Gamma(1-\alpha-\epsilon)\over
  (4\pi)^{1-\epsilon}\Gamma(\alpha)} \left(b\over 2\right)^{2\alpha-2+2\epsilon}\,,
\end{equation}
the expansion in eq.~\eqref{e:genericExpansion} reads
\begin{multline}\label{e:genericExpansionb}
  \widetilde{\mathcal M^{N=8}_L}(\phi,\phi,\phi,\phi)=   {1\over \pi
    \hbar^{L-1}b^2}\left(b^{2(L+1)}\pi\over 4^L\right)^\epsilon 
\frac{4m_1^3m_2^3\sigma^4}{\sqrt{\sigma^2-1}}\cr
 \times \sum_{r\geq-2}  \mathcal M_L^{(r)}(\sigma,\epsilon)\left(2\hbar\over b\right)^r
 \frac{ \Gamma \left(\frac{r}{2}-L \epsilon -\epsilon +1\right)}{\Gamma \left(L \epsilon -\frac{r}{2}\right)} .
\end{multline}
For  the choice of helicity
in~\cite{DiVecchia:2021ndb,DiVecchia:2020ymx}  with the kinematic
factor in eq.~\eqref{e:KV}  we have
\begin{equation}\label{e:MbHelV}
\widetilde{\mathcal M^{N=8}_L}(\phi,\phi,\varphi,\varphi)\!=\!\frac{4m_1^3m_2^3\sigma^4}{\sqrt{\sigma^2-1}}\!\!\int\!\!\!
\frac{d^{D-2}\vec{\underline q}}{(2\pi)^{D-2}}\!\left(\!1\!-\!{\hbar^2 \underline
    q^2\over m_1m_2\sigma}+{\cal O}(\hbar \underline q)^3\!\right)\mathcal M_L(p_1,p_2,p_1',p_2') e^{i \vec{\underline q}\cdot\vec{b}}\,.
\end{equation}
We will show that the choice of helicity does not affect the classical
piece of the amplitude.

\subsubsection{Tree-level amplitude}
Using the expansion given previously we have for the tree amplitude
\begin{equation}
  \label{e:Mtreephi4b}
  \widetilde{\mathcal M^{N=8}_0}(\phi,\phi,\phi,\phi)= {1\over \hbar}
   \frac{2m_1m_2\sigma^2G_N}{\sqrt{\sigma^2-1}} (b\sqrt\pi)^{2\epsilon}\Gamma(-\epsilon)\,.
 \end{equation}
We notice that this expression is actually exact because any higher
order terms in the $|{\underline q}|^2$ expansion in the
tree-level amplitude has a vanishing Fourier transform.\\[5pt]
 As a consequence of the vanishing of any higher power correction in
 $|{\underline q}|^2$, 
we have for the helicity choice in~\cite{DiVecchia:2021ndb,DiVecchia:2020ymx}  with the kinematic
factor in eq.~\eqref{e:KV} 
 \begin{equation}
  \label{e:MtreephiV4b}
  \widetilde{\mathcal M^{N=8}_0}(\phi,\phi,\varphi,\varphi)= \widetilde{\mathcal M^{N=8}_0}(\phi,\phi,\phi,\phi)\,.
\end{equation}

 \subsubsection{One-loop amplitude}
For the one-loop amplitude  to all order in $\epsilon$
  \begin{equation} \label{e:Monephi4b}
     \widetilde{\mathcal M^{N=8}_1}(\phi,\phi,\phi,\phi)={i\over2} \left(  \widetilde{\mathcal
      M^{N=8}_0}(\phi,\phi,\phi,\phi)\right)^2+  \widetilde{\mathcal
    M^{N=8}_1}(\phi,\phi,\phi,\phi)|_{\rm Cl.}+  \widetilde{\mathcal
    M^{N=8}_1}(\phi,\phi,\phi,\phi)|_{\rm Qt}\,.
\end{equation}
Where the one-loop classical part is given by
\begin{align}\label{e:Moneloopclassical}
     \widetilde{\mathcal M^{N=8}_1}(\phi,\phi,\phi,\phi)|_{\rm
       Cl.}&=
      {1\over 4\pi}
  \left(b^2 \sqrt\pi \over2\right)^{2\epsilon}\frac{4m_1^3m_2^3\sigma^4}{\sqrt{\sigma^2-1}} {2 \mathcal
    M_1^{(-1)}(\sigma,\epsilon)\over \hbar
                   b}{\Gamma(\frac12-2\epsilon)\over\Gamma(\frac12+\epsilon)}\displaystyle\\
\nonumber                   &=\frac{4 G_N^2 m_1 m_2 \sigma ^4 
  (m_1+m_2) }{ \left(\sigma
   ^2-1\right)^{3\over2}\hbar }  { \left(b^2\pi\right)^{2 \epsilon -\frac12} \Gamma \left(\frac{1}{2}-2
   \epsilon \right) \Gamma \left(\frac{1}{2}-\epsilon \right)^2\over\Gamma (-2 \epsilon ) }\displaystyle\,,
\end{align}
and the quantum piece of the one-loop amplitude is given by
\begin{align}\label{e:Moneloopquantum}\displaystyle
     \widetilde{\mathcal M^{N=8}_1}(\phi,\phi,\phi,\phi)|_{\rm Qt.}&= 
{1\over 4\pi}
  \left(b^2 \sqrt\pi \over2\right)^{2\epsilon}\frac{4m_1^3m_2^3\sigma^4}{\sqrt{\sigma^2-1}}
  {4\mathcal M_1^{(0)}(\sigma,\epsilon)\over b^2}
  {\Gamma(1-2\epsilon)\over \Gamma(\epsilon)}\\
  &=-\frac{4 i G_N^2 m_1 m_2 \sigma ^4 \displaystyle
   \left(\frac{\pi  s \epsilon }{2 m_1 m_2}+i (2 \epsilon +1) \left(\sigma 
   \arccosh(\sigma )-\sqrt{\sigma ^2-1}\right)\right)}{\left(\sigma ^2-1\right)^2
    } \cr
\nonumber    &\times{ \left(b^2\pi\right)^{2
   \epsilon -1} \Gamma (1-2 \epsilon ) \Gamma (-\epsilon )^2 \Gamma (\epsilon +1) \over \Gamma (-2 \epsilon ) \Gamma (\epsilon )}\displaystyle\,.
\end{align}
For  the choice of helicity
in~\cite{DiVecchia:2021ndb,DiVecchia:2020ymx}  with the kinematic
factor in eq.~\eqref{e:KV} 
 \begin{equation} \label{e:MonephiV4b}
     \widetilde{\mathcal
       M^{N=8}_1}(\phi,\phi,\varphi,\varphi)=\widetilde{\mathcal
       M^{N=8}_1}(\phi,\phi,\phi,\phi)- {\mathcal
     M_1^{(-2)} (\sigma,\epsilon) \over b^2 m_1m_2\pi\sigma}\left(b^4 \pi\over
       4\right)^\epsilon{\Gamma(1-2\epsilon)\over\Gamma(\epsilon)}
  \,.
\end{equation}
The effect of the helicity choice on the only amplitude only affects
the quantum part of the amplitude not the classical part so that
\begin{align}\label{e:M1helrelation}
\widetilde{\mathcal
  M^{N=8}_1}(\phi,\phi,\varphi,\varphi)|_{\rm Cl.}&=    \widetilde{\mathcal
  M^{N=8}_1}(\phi,\phi,\phi,\phi)|_{\rm Cl.},\\
  \widetilde{\mathcal
       M^{N=8}_1}(\phi,\phi,\varphi,\varphi)|_{\rm Qt.}&=    \widetilde{\mathcal
  M^{N=8}_1}(\phi,\phi,\phi,\phi)|_{\rm Qt.}- { \mathcal
     M_1^{(-2)}(\sigma,\epsilon)\over b^2 m_1m_2\pi\sigma} \left(b^4 \pi\over
       4\right)^\epsilon {\Gamma(1-2\epsilon)\over\Gamma(\epsilon)}\,.
\end{align}
%
\subsubsection{Two-loop amplitude}\label{sec:twoloop}
For the two-loop amplitude
\begin{multline}
  \label{e:Mtwophi4b}
  \widetilde{\mathcal M^{N=8}_2}(\phi,\phi,\phi,\phi)=-{1\over 6} \left(  \widetilde{\mathcal
      M^{N=8}_0}(\phi,\phi,\phi,\phi)\right)^3
  +i \widetilde{\mathcal
      M^{N=8}_0}(\phi,\phi,\phi,\phi) \widetilde{\mathcal
      M^{N=8}_1}(\phi,\phi,\phi,\phi) |_{\rm Cl.}\cr
    +i \widetilde{\mathcal M^{N=8}_0}(\phi,\phi,\phi,\phi)|\widetilde{\mathcal M^{N=8}_1}(\phi,\phi,\phi,\phi)|_{\rm
    Qt.}
+  \widetilde{\mathcal M^{N=8}_2}(\phi,\phi,\phi,\phi)|_{\rm Cl.},
\end{multline}
with the classical piece at two-loop given by
\begin{multline}
  \label{e:Mtwophi4bclassical}\displaystyle
  \widetilde{\mathcal M^{N=8}_2}(\phi,\phi,\phi,\phi)|_{\rm
    Cl.}=\frac{16G_N^3m_1^2 m_2^2\sigma^4(\pi b^2 e^{\gamma_E})^{3\epsilon}}{\hbar b^2 (\sigma^2-1)^{\frac52}} \Bigg[ -(\sigma^2-1)^{\frac32}\arccosh(\sigma)\\- \frac{i}{\pi \epsilon}\bigg(\frac{-1}{4(\sigma^2-1)} \bigg)^{\epsilon} \Bigg( (1+2\epsilon)\sigma^2\sqrt{\sigma^2-1}+\sigma(\sigma^2-2)\arccosh(\sigma)\\+\epsilon \big((\sigma^2-1)^{\frac32}-\sigma(\sigma^2-2)\big) \arccosh^2(\sigma) -\epsilon \sigma(\sigma^2-2) \Li_2\left(2-2\sigma(\sigma+\sqrt{\sigma^2-1})\right) \Bigg) +{\cal O}(\epsilon) \Bigg]\,.
\end{multline}
The contribution $ i \widetilde{\mathcal M^{N=8}_0}(\phi,\phi,\phi,\phi)|\widetilde{\mathcal M^{N=8}_1}(\phi,\phi,\phi,\phi)|_{\rm
    Qt.}$ is of $1/\hbar$ order as the classical contribution. But
  we have separated it off from the classical contribution because
  only eq.~\eqref{e:Mtwophi4bclassical} will contribute to the  classical
  scattering angle, as will be shown in section~\ref{sec:scattering-angle}.\\[5pt]
For  the choice of helicity
in~\cite{DiVecchia:2021ndb,DiVecchia:2020ymx}  with the kinematic
factor in eq.~\eqref{e:KV} 
we have 
\begin{multline}
  \label{e:Mtwophi4bV}
  \widetilde{\mathcal
    M^{N=8}_2}(\phi,\phi,\varphi,\varphi)=\widetilde{\mathcal
    M^{N=8}_2}(\phi,\phi,\phi,\phi)-\left(b^6\pi \over
    16\right)^\epsilon {\mathcal M_2^{(-2)}(\sigma)\over
  \hbar b^2 m_1m_2\pi\sigma}{\Gamma(1-3\epsilon)\over\Gamma(2\epsilon)}.
\end{multline}
The extra classical piece generated by the $|{\underline q}|^2$
term in the kinematic factor in eq.~\eqref{e:KV} goes into the
modification of the quantum one-loop part as given
in eq.~\eqref{e:M1helrelation} so that the two-loop amplitude is
decomposed as
\begin{multline}
  \label{e:Mtwophi4bVres}
  \widetilde{\mathcal M^{N=8}_2}(\phi,\phi,\varphi,\varphi)=-{1\over 6} \left(  \widetilde{\mathcal
      M^{N=8}_0}(\phi,\phi,\varphi,\varphi)\right)^3
  + i \widetilde{\mathcal
      M^{N=8}_0}(\phi,\phi,\phi,\phi) \widetilde{\mathcal
      M^{N=8}_1}(\phi,\phi,\varphi,\varphi) |_{\rm Cl.}\cr
    +i \widetilde{\mathcal M^{N=8}_0}(\phi,\phi,\phi,\phi)|\widetilde{\mathcal M^{N=8}_1}(\phi,\phi,\varphi,\varphi)|_{\rm
    Qt.}
+  \widetilde{\mathcal M^{N=8}_2}(\phi,\phi,\varphi,\varphi)|_{\rm Cl.},
\end{multline}
with
\begin{equation}
     \widetilde{\mathcal M^{N=8}_2}(\phi,\phi,\varphi,\varphi)|_{\rm Cl.}= \widetilde{\mathcal M^{N=8}_2}(\phi,\phi,\phi,\phi)|_{\rm Cl.}\,.
\end{equation}
Since the classical part of the two-loop amplitude does not depend on
the helicity of the external state we just set
\begin{equation}
     \mathcal M^{\rm Cl.}_2(\sigma,b)\equiv \widetilde{\mathcal M^{N=8}_2}(\phi,\phi,\varphi,\varphi)|_{\rm Cl.}.
\end{equation}
The classical contribution to the two-loop amplitude is composed of a real and imaginary part
\begin{equation}
  \mathcal M_2^{\rm Cl.}(\sigma,b)= \mathcal M_2^{\rm Cl.~R}(\sigma,b)+ 
i \mathcal M_2^{\rm Cl.~I}(\sigma,b)+{\cal O}(1).
\end{equation}
\paragraph{The real part of  the two-loop classical amplitude} is free of 
divergences
\begin{equation}
  \mathcal M_2^{\rm Cl.~R}(\sigma,b)=\Re(\mathcal M_2^{\rm Cl.}(\sigma,b)) = 
                         \mathcal M_2^{\rm Cl.~R}(\sigma,b)|_{\rm Cons.}+
  \mathcal M_2^{\rm Cl.~R}(\sigma,b)|_{\rm Rad.}\,,
\end{equation}
and composed of a conservative part
 \begin{equation}
\mathcal M_2^{\rm Cl.~R} (\sigma,b)|_{\rm Cons.}    = 
 - \frac{8 G_N^3 m_1^2 m_2^2 \sigma^4 }{b^2}
   {\arccosh(\sigma )\over \sigma^2-1}+{\cal O}(\epsilon)\,,
\end{equation}
a radiation-reaction part  
  \begin{equation}
\mathcal M_2^{\rm Cl.~R}(\sigma,b)|_{\rm Rad.} \!=\! 
  \frac{8 G_N^3 m_1^2 m_2^2 \sigma^4 }{b^2  (\sigma^2-1)^\epsilon}
 \left(\pi b^2e^{\gamma_E}\over4\right)^\epsilon
   \!\! \left(    {\sigma^2 \over (\sigma ^2-1)^{2}}+  {\sigma(\sigma ^2-2)\over (\sigma
   ^2-1)^{{5\over2}}}\arccosh(\sigma )\!\!\right)\!+\!{\cal O}(\epsilon)\,.
\end{equation}

\paragraph{The imaginary part of  the scattering phase}
is given by 
\begin{align}\label{e:genericdelta2}
  \mathcal M_2^{\rm Cl.~I}(\sigma,b)&=\Im(\mathcal M_2^{\rm Cl.}(\sigma,b)) \\
  &= -{\mathcal M_2^{\rm Cl.~R}(\sigma,b)|_{\rm Rad.} \over\pi\epsilon}+
  \frac{8 G_N^3 m_1^2 m_2^2 \sigma^4 }{ \pi b^2(\sigma^2-1)^{\epsilon}}\left(\pi b^2e^{\gamma_E}\over4\right)^\epsilon\cr
&\times                                                                                                                                         
                     \Bigg[-{2\sigma \over(\sigma ^2-1)^2 }+{\sigma ^2-2\over(\sigma ^2-1)^{5\over2}}
 \Li_2\left(2-2\sigma(\sigma+\sqrt{\sigma^2-1})\right) \cr
\nonumber &+\left({1 \over \sigma(\sigma ^2-1)}+ {\sigma ^2-2
  \over (\sigma ^2-1)^{5\over2}}\right) \arccosh(\sigma
            )^2\Bigg]+{\cal O}(\epsilon)\,.
\end{align}
This result establishes the relation  conjectured
in~\cite{DiVecchia:2021ndb} between the radiation-reaction part of the real
part and the infrared divergence of  imaginary part
\begin{equation}
  \lim_{\epsilon\to0} \mathcal M_2^{\rm Cl.~R}(\sigma,b)|_{\rm Rad.}=-\lim_{\epsilon\to0} \epsilon\pi  \mathcal M_2^{\rm Cl.~I}(\sigma,b)\,.
\end{equation}

\paragraph{At high-energy $\sigma\gg1$} we have that
\begin{multline}
    -{2\sigma \over(\sigma ^2-1)^2 }+{\sigma ^2-2\over(\sigma ^2-1)^{5\over2}}
    \Li_2\left(2-2\sigma(\sigma+\sqrt{\sigma^2-1})\right) \cr
    +\left({1 \over
     \sigma(\sigma ^2-1)}+ {\sigma ^2-2
  \over (\sigma ^2-1)^{5\over2}}\right) \arccosh(\sigma )^2=
  -{\zeta(2)+2\over \sigma^3}+\mathcal O(\sigma^{-2}),
 \end{multline}
 and
 \begin{equation}
     {\sigma \over (\sigma ^2-1)^{2}}+  {\sigma ^2-2\over (\sigma
   ^2-1)^{{5\over2}}}\arccosh(\sigma )={1+\log(2\sigma)\over
   \sigma^3}+\mathcal O(\sigma^{-2}),
 \end{equation}
 therefore by expanding the factor of
 $(\sigma^2-1)^{-\epsilon}=1-\epsilon\log(\sigma^2-1)$ 
we have that the radiation part reads  for $\sigma\gg1$
\begin{equation}
  \lim_{\sigma\gg1} \mathcal M_2^{\rm Cl.}(\sigma,b)\simeq  \left(1+{i\over
      \pi}\left(-{1\over\epsilon}+ \log(\sigma^2-1)\right)\right) \textrm{coeff}_{\epsilon^0}(\mathcal M_2^{\rm Cl.~R}(\sigma,b)|_{\rm Rad.}).
\end{equation}
This shows that the $\log(\sigma^2-1)$ of eq.~(3.2) in~\cite{DiVecchia:2021ndb}
arises from the $\epsilon$ expansion of the
$(\sigma^2-1)^{-\epsilon}$ in eq.~\eqref{e:genericdelta2}.

\paragraph{The soft and potential region.}
We now consider the limits $\sigma\to1$ and $\epsilon\to0$ of the
result.

\noindent{$\bullet$}
In the limit $\epsilon\to0$ with $\sigma$ fixed
\begin{equation}
\lim_{\epsilon\to0\atop
  \sigma~\textrm{fixed}} \Re(\mathcal M_2^{\rm Cl.})=  \frac{8\pi G_N^3 
m_1^2 m_2^2 \sigma^4 }{b^2}
\left({\sigma^2\over (\sigma ^2-1)^2}+\arccosh(\sigma ) \left(\sigma{\sigma ^2-2\over(\sigma
   ^2-1)^{5\over2}}-{1\over\sigma^2-1}\right)\right),
\end{equation}
we recover the result from the soft region~\cite{DiVecchia:2021bdo}.

\noindent{$\bullet$}
Whereas in the  the $\sigma\to1$ limit with $\epsilon<0$ fixed
we have
\begin{equation}
  \lim_{\sigma\to 1\atop
  \epsilon<0~\textrm{fixed}} \Re(\mathcal M_2^{\rm Cl.})=\frac{8 G_N^3 m_1^2 m_2^2 \sigma^4}{b^2}\left(-{ \arccosh(\sigma)\over (\sigma^2-1)}+
            {5\over 3(\sigma^2-1)}+{\cal O}(\sigma-1)^0\right)
              +{\cal O}(\epsilon).
\end{equation}
The leading contribution when $\sigma\to1$ is the conservative piece computed
in~\cite{Parra-Martinez:2020dzs}  whereas the sub-leading piece is the 
radiation-reaction contribution~\cite{Damour:2020tta,DiVecchia:2021ndb}.

\subsection{The eikonal phase}
\label{sec:scattering-angle}
The  full scattering matrix in
the $b$-space can be expanded as~\cite{DiVecchia:2019kta}
\begin{equation}
1+i  \mathcal T (\lambda_1,\dots,\lambda_4) =1+i\sum_{L\geq0}\widetilde{\mathcal M^{N=8}_L}(\lambda_1,\dots,\lambda_4)=(1+i2\Delta (\lambda_1,\dots,\lambda_4)) e^{i2
    \delta(\sigma,b)\over \hbar}  \displaystyle\,,
\end{equation}
where $\widetilde{\mathcal
  M^{N=8}_L}(\lambda_1,\dots,\lambda_4)$ are the $L$-loop amplitudes,  $\delta$ is the classical eikonal and $\Delta$ a quantum
correction.\\[5pt]
In perturbative expansion we have 
\begin{equation}
  \delta= \delta_0+\delta_1+\delta_2+\cdots; \qquad \Delta=\Delta_1+\Delta_2+\cdots\,,
\end{equation}
with $\delta_r$ and $\Delta_r$  are  of order $G_N^{r+1}$, 
gives to the first order studied explicitly in this work
\begin{align}
  2\delta_L&=\hbar\Re(\widetilde{\mathcal M_L}|_{\rm Cl.}),  \qquad L=0,1,2\cr
        2    \Delta_1&=  \widetilde{  \mathcal M_1}|_{\rm Qt.}   \cr
2\Delta_2&=-i      \widetilde{ \mathcal M_0}        \widetilde{    \mathcal M_1} |_{\rm Qt.}+\Im(\widetilde{\mathcal M_2}|_{\rm Cl.}) \,.
\end{align}
The exponentiation 
of the perturbative expansion $1+i \sum_L \mathcal
M_L(\sigma,b)$ is only possible if the partial amplitudes
$\mathcal M_L^{(r)}(\sigma,\epsilon)$ satisfy relations similar to
the one noticed   at one-loop in eq.~\eqref{e:Monephi4b}
and two-loop in eq.~\eqref{e:Mtwophi4b}.
In particular the exponentiation is possible because the
contributions more singular than the classical one in the $\hbar\to0$
limit do satisfy unitarity related  relations in 
$b$-space~\cite{Cristofoli:2020uzm}.  \\[5pt]
We have noticed before that the different choices of helicity made
in~\cite{Parra-Martinez:2020dzs} and~\cite{DiVecchia:2021ndb,DiVecchia:2020ymx} do not affect the
classical part of the eikonal phase, and therefore $\delta$ is the same for
all helicity choices as expected from the universality of  classical
gravitational interactions. Only
the quantum part $\Delta$ depends on the helicity choice.  A similar dependence helicity
dependence of the external states on the
quantum part of the one-loop amplitude was noticed
in~\cite{Bjerrum-Bohr:2014zsa}.\\[5pt]
At tree-level $L=0$ and one-loop $L=1$ orders the classical piece of
the amplitude is real, and the eikonal phase is equal to the
classical part of the amplitude. The classical part of the two-loop $L=2$ amplitude has an imaginary
part (that was discussed above). We have decided to only exponentiate
the real part of the classical two-loop amplitude since an imaginary
contribution to the eikonal phase would violate unitarity. But this would 
have to
be confirmed by an higher loop computation.\\[5pt]
Therefore, the
classical scattering angle is obtained by the stationary
phase  from the (real)  eikonal phase 
\begin{equation}\label{e:chidef}
\sin\left(\frac{\chi}{2}\right)=-{\sqrt{s}\over m_1 m_2\sqrt{\sigma^2-1}}\frac{\partial \delta(\sigma,b)}{\partial b}\,,
\end{equation}
and does not depend on the helicity choice. Universality of the high-energy limit of gravitational scattering  up to two-loop order was first demonstrated in
ref.~\cite{Bern:2020gjj}.

\subsubsection{The first Post-Minkowskian order}
Because the  tree-level amplitude is independent of the helicity
choice the scattering phase is given by the
$2\delta_0=\hbar\times\eqref{e:Mtreephi4b}$ 
therefore the scattering angle at the first Post-Minkowskian order is given by
\begin{align}
\left.\sin\left(\frac{\chi}{2}\right)\right|_{1PM}&=\frac{2G_N \sqrt{s}
                                         \sigma^2 }{b(\sigma^2-1)}  (b\sqrt\pi)^{2\epsilon}\Gamma(1-\epsilon)\cr
                                         &= \frac{2G_N \sqrt{s}
                                         \sigma^2 }{b(\sigma^2-1)}+{\cal O}(\epsilon)\,.
\end{align}
%
\subsubsection{The second Post-Minkowskian order}
Because the helicity choice does not affect the classical piece of the
one-loop amplitude, 
the one-loop scattering phase is given by
$2\delta_1=\hbar\times\eqref{e:Moneloopclassical}$ leading to the
scattering angle 
\begin{align}
\left.\sin\left(\frac{\chi}{2}\right)\right|_{2PM}&=\frac{8G_N^2 \sqrt{s} \sigma ^4 
  (m_1+m_2) }{ \left(\sigma
   ^2-1\right)^2  b}{ \left(2b^2\pi\right)^{2 \epsilon}\Gamma \left(\frac{3}{2}-2
   \epsilon \right)\Gamma \left(\frac{1}{2}-\epsilon \right)\over
                                         \Gamma (-\epsilon )}\cr
                                         &=0+{\cal O}(\epsilon)\,.
\end{align}
At the leading order in $\epsilon$ the scattering angle at the second Post-Minkowskian order vanishes.
A result which has been linked in~\cite{Caron-Huot:2018ape} to  the no-triangle property of maximal
supergravity amplitude~\cite{BjerrumBohr:2008ji}.\\[5pt]
The $\Delta_1$  is given by the leading quantum
part of the one-loop amplitude. The  result for the helicity choice
in~\cite{Parra-Martinez:2020dzs}  and in~\cite{DiVecchia:2021ndb,DiVecchia:2020ymx} differ by the shift in eq.~\eqref{e:MonephiV4b}.
%
\subsubsection{The third Post-Minkowskian order}
The two-loop order scattering phase is given by $2\delta_2=\hbar
\times\Re\eqref{e:Mtwophi4bclassical}$ which does not depend on the
helicity configuration as explained before
\begin{multline}
  \delta_2(\sigma,b)=  \Re\Bigg(\frac{8G_N^3m_1^2 m_2^2\sigma^4(\pi b^2 
e^{\gamma_E})^{3\epsilon}}{b^2 (\sigma^2-1)^{\frac52}} \Bigg[
-(\sigma^2-1)^{\frac32}\arccosh(\sigma)\\- \frac{i}{\pi
  \epsilon}\bigg(\frac{-1}{4(\sigma^2-1)} \bigg)^{\epsilon} \Bigg(
(1+2\epsilon)\sigma^2\sqrt{\sigma^2-1}+\sigma(\sigma^2-2)\arccosh(\sigma)\\+\epsilon
\big((\sigma^2-1)^{\frac32}-\sigma(\sigma^2-2)\big) \arccosh^2(\sigma)
-\epsilon \sigma(\sigma^2-2)
\Li_2\left(2-2\sigma(\sigma+\sqrt{\sigma^2-1})\right)\Bigg) +{\cal O}(\epsilon) \Bigg]\Bigg)\,.
\end{multline}
The scattering angle at the third Post-Minkowskian order is then
given by
\begin{equation}
     \left.\sin\left(\chi\over2\right)\right|_{3PM}=-{\sqrt{s}\over m_1m_2\sqrt{\sigma^2-1}}{\partial \delta_2(\sigma,b)\over \partial b}\,,
   \end{equation}
   and reads
\begin{multline}
  \label{e:chi3PM}
  \left.\sin\left(\chi\over2\right)\right|_{3PM}=\frac{16 G_N^3
    \sqrt{s} m_1
    m_2 \sigma^4 }{b^3
    (\sigma^2-1)}\Bigg(-\frac{\arccosh(\sigma)}{\sqrt{\sigma^2-1}}\cr
  +{1\over
      (4(\sigma^2-1))^{\epsilon}}\left(\frac{\sigma(\sigma^2-2) \arccosh(\sigma)}{(\sigma^2-1)^{2}}+\frac{\sigma^2}{(\sigma^2-1)^{\frac{3}{2}}}\right)\Bigg)+{\cal O}(\epsilon),
  \end{multline}
which presents a conservative part in the first line and  a radiation-reaction
part, given in the second line.\\[5pt]
We will see below how the $\epsilon$-expansion of the the soft factor $(\sigma^2-1)^{-\epsilon}$  matches with the results
of~\cite{DiVecchia:2021ndb}. \\[5pt]
Using the
angular momentum
\begin{equation}\label{e:Jdef}
  J={m_1 m_2\sqrt{\sigma^2-1}\over \sqrt{s}}b \cos(\frac{\chi}{2})\,,
  \end{equation}
we  can decompose  the scattering angle at the third Post-Minkowskian
order 
\begin{equation}
\chi_{3PM}=\chi_{3PM}^{\rm Schw.}+\chi^{\rm
Cons.}_{3PM}+\chi_{3PM}^{\rm Rad.}\,,
\end{equation}
into a  Schwarzschild metric contribution
\begin{equation}
    \chi_{3PM}^{\rm Schw.}=-16m_1^3 m_2^3 G_N^3\frac{ \sigma^6 }{3J^3(\sigma^2-1)^{\frac{3}{2}}},
  \end{equation}
  a conservative part
  \begin{equation}
\label{e:ChiCons}    
  {\chi}_{3PM}^{\rm Cons.}=-32 m_1^4 m_2^4 \sigma^4 G_N^3\frac{\arccosh(\sigma)}{J^3 s}\,,
\end{equation}
and a radiation part
\begin{equation}
  \label{e:chiRad}
  {\chi}_{3PM}^{\rm  Rad.}=32 m_1^4 m_2^4 G_N^3 \frac{\sigma^4}{J^3
    s}{1\over 
   (4 (\sigma^2-1))^{\epsilon}}\left(\frac{\sigma(\sigma^2-2) \arccosh(\sigma)}{(\sigma^2-1)^{\frac{3}{2}}}+\frac{\sigma^2}{\sigma^2-1}\right)\,.
\end{equation}
When setting $(\sigma^2-1)^{-\epsilon}=1$ these results  reproduce
eq~(4.7) of~\cite{DiVecchia:2020ymx}. We remark that taking the limit $\sigma\to1$
with  fixed $J$,  gives  $ {\chi}_{3PM}^{\rm  Rad.}=0$ because $\epsilon<0$.
%
\section{The evaluation of the two-loop scalar double boxes}\label{sec:evaluationdoubleboxes}
In this section we provide details about the evaluation of the
double-box leading to the results in section~\ref{sec:twoloopresults}. \\[5pt]
The two-loop amplitude in eq.~\eqref{e:Mtwoloop} after the Kaluza-Klein
reduction  in eq.~\eqref{e:KKreduc} reads
\begin{equation}\label{e:M2loop}
 \mathcal{M}^{ 2-{\rm loop}}(p_1,p_2,p_1',p_2')=(8 \pi G_N)^3
   \left(4 m_1^2m_2^2\sigma^2 (J_s\!+\!J_u)\!+\!2\hbar^2 m_1m_2
 | \underline{\vec q}|^2\sigma
J_u \!+\! \hbar^4 |\underline q|^4 J_t
\right)\,.
\end{equation}
We see that the classical contribution arises from the
contribution of order  $1/\hbar$ from $J_s+J_u$, from the contribution
of order $1/\hbar^3$ from $J_u$ and from the contribution of order
$1/\hbar^5$ from $J_t$.\\[5pt]
We first give a summary of the results for the expansion of these
contributions and discuss the comparison with the existing results in
the literature. We will then gives details on the methods used for
deriving these results.

\subsection{The $t$-channel contribution}
The $t$-channel contributions are given by the so-called $H$ diagrams
integrals. Using the $q=|q|u_q$ and $q=\hbar \underline q$, by
rescaling the loop integrations $\ell_i\to l_i |q|$ we have
\begin{multline}
\!\!\!J_t=\frac{-|\underline q|^{2D-12}}{2\hbar^5}\!\!\int\! \frac{d^D l_1
  d^D l_2}{(2\pi)^{2D}}\! \left(\!\frac{1}{2p_1\cdot l_1+i
  \varepsilon}-\frac{1}{2{p_1'}\cdot l_1-i \varepsilon}\!\right)\! \left(\!\frac{1}{2p_2\cdot
  l_2-i \varepsilon}\!-\!\frac{1}{2{p_2'}\cdot l_2+i \varepsilon}\!\right)\cr
\times
\frac{1}{((l_1+u_q)^2+i\varepsilon) ((l_2+u_q)^2+i\varepsilon)
  (l_1^2+i\varepsilon)
  (l_2^2+i\varepsilon) ((l_1+l_2+u_q)^2+i\varepsilon)}.
\end{multline}
Because the integral is already of the order $1/\hbar^5$ is it enough
to keep the leading order in the $|\underline q|$ expansion for
extracting the classical piece\footnote{The power of $\hbar$ arises
  because this diagram as two massive propagators and five massless
  propagators giving  a power of $\hbar$, to be
  $3L+1-(2+2(3L-1))=1-3L=-5$ for $L=2$. But the total amplitude in eq.~\eqref{e:M2loop} has a
  power $\hbar^4$ in the  numerator. }
\begin{equation}
  J_t=-\frac{|\underline q|^{-4-4\epsilon}}{16 m_1 m_2\hbar^5\epsilon^4
  \sqrt{\sigma^2-1}} \, \mathcal I_6(\sigma)+{\cal O}(\hbar |\underline q|)\,,
\end{equation}
where $\mathcal I_6(\sigma)$ is one of the two-loop master
integrals\footnote{In  appendix~\ref{sec:deltatomaster} we explain how
  to convert the double-box integrals to the
  integrals with a generalized propagator that is used in the
  definition of the master integrals.} which is evaluated in section~\ref{e:I6evaluation} the final result is
\begin{multline}\label{e:Jtresult}
J_{t}=-\frac{1}{256 m_1 m_2\hbar^5\pi^3 \epsilon}{ (4\pi
e^{- \gamma_E})^{2\epsilon}\over  |\underline q|^{4+4\epsilon}} {\arccosh\left(\sigma\right)\over
\sqrt{\sigma^2-1}}\cr
\times\left(\pi+i \Big( \frac{-1}{4(\sigma^2-1)}\Big)^{\epsilon}
  \arccosh\left(\sigma\right)+{\cal O}(\epsilon)
\right)+\mathcal O(\hbar |\underline q|)\,.
\end{multline}
%

\subsection{The $u$-channel contribution}
For the $u$-channel contribution we have
\begin{multline}
  J_u= {|\underline q|^{2D-10}\over 96\hbar^3} \int
  {d^Dl_1d^Dl_2\over(2\pi)^D} \frac{1}{(l_1^2+i\varepsilon) (l_2^2+i\varepsilon) ((l_1+l_2+u_q)^2+i\varepsilon)}\cr\sum_{1\leq i\neq j\leq 3\atop  
 i\neq
    k,j\neq n}\sum_{1\leq k\neq n\leq 3} \frac{1}{(\bar{p}_1 \cdot
    l_i+|\vec{q}| \frac{u_q \cdot l_i}{2}+i \varepsilon)(\bar{p}_1
    \cdot l_j-|\vec{q}| \frac{u_q \cdot l_j}{2}-i \varepsilon)}\cr
  \times\frac{1}{(\bar{p}_2 \cdot l_k-|\vec{q}| \frac{u_q \cdot l_k}{2}-i \varepsilon)(\bar{p}_2 \cdot l_n+|\vec{q}| \frac{u_q \cdot l_n}{2}+i \varepsilon)}   \,.
\end{multline}
We introduced the momenta $\bar{p_i}=p_1-q/2$ and $\bar p_2=p_2+q/2$
which are orthogonal to $q$. And as before we rescaled the loop
momenta by $\ell_i\to l_i |q|$ with $q=u_q |q|$,  and $q=\hbar \underline q$,
\begin{multline}
   J_{u }^0=  {|\underline q|^{2D-10}\over 96\hbar^3} \int
  {d^Dl_1d^Dl_2\over(2\pi)^D} \frac{1}{(l_1^2+i\varepsilon) (l_2^2+i\varepsilon) ((l_1+l_2+u_q)^2+i\varepsilon)}\cr\sum_{1\leq i\neq j\leq 3\atop  
 i\neq
    k,j\neq n}\sum_{1\leq k\neq n\leq 3} {1\over (\bar p_1\cdot
    l_i+i\varepsilon)(\bar p_1\cdot l_j-i\varepsilon)(\bar p_2\cdot
    l_k-i\varepsilon)(\bar p_2\cdot l_n+i\varepsilon)} \,.
\end{multline}
Using the  principal part formula in eq.~\eqref{e:PP}, we can rewrite this expression
using delta functions
\begin{multline}
J_{u }^0=\frac{|\underline q|^{2D-10}}{192\hbar^3} \int \frac{d^D l_1
  d^D l_2}{(2\pi)^{2D-4}} \frac{\delta(\bar{p}_1 \cdot
  l_1)\delta(\bar{p}_1 \cdot l_2)\delta(\bar{p}_2 \cdot
  l_1)\delta(\bar{p}_2 \cdot l_2)}{l_1^2 l_2^2 (l_1+l_2+u_q)^2}\\
-\frac{|\underline q|^{2D-10}}{64\hbar^3}\int \frac{d^D l_1 d^D
  l_2}{(2\pi)^{2D-2}}  \frac{\delta(\bar{p}_1 \cdot l_1)
  \delta(\bar{p}_2 \cdot l_2)}{l_1^2 l_2^2
  (l_1+l_2+u_q)^2}\left(\frac{1}{\bar{p}_1\cdot l_2-i\varepsilon}+\frac{1}{\bar{p}_1\cdot
  l_2+i\varepsilon}\right)\left(\frac{1}{\bar{p}_2\cdot
  l_1-i\varepsilon}+\frac{1}{\bar{p}_2\cdot l_1+i\varepsilon}\right)\,,
\end{multline}
which can be expressed using the master integral $\mathcal I_9(\sigma)$ as
\begin{equation}
  J_u^0=\frac{|\underline q|^{-2-4\epsilon}}{16\hbar^3
    m_1^2m_2^2(\sigma^2-1)\epsilon^4} (\mathcal I_9(\sigma)-b_9 \epsilon^2)\,.
\end{equation}
Because
the integral is of order $1/\hbar^3$ one needs the leading order
expansion of the the integral for
getting the  classical contribution
\begin{equation}
    J_{u }=\frac{|\underline q|^{-2-4\epsilon}}{16\hbar^3
    m_1^2m_2^2(\sigma^2-1)\epsilon^4} (\mathcal I_9(\sigma)-b_9
  \epsilon^2)+\mathcal O(\hbar|\underline q|)\,,
\end{equation}
where $\mathcal I_9(\sigma)$ is one of the two-loop master integrals
evaluated in section~\ref{sec:I9evaluation}
and $b_9$ is the constant of integration determined in eq.~\eqref{e:b9}.
At the leading order in $|\underline q|$ in $D=4-2\epsilon$ we get that 

\begin{multline}\label{e:Ju0results}
    J_u=   \frac{i(4\pi e^{-\gamma_E})^{2\epsilon}}{512\pi^3\hbar^3
    m_1^2m_2^2 |\underline q|^{2+4\epsilon}(\sigma^2-1)\epsilon^2} \Big(\frac{-1}{4(\sigma^2-1)}\Big)^{\epsilon}
 \cr
  \times\left(\arccosh(\sigma)-\epsilon
    \left(\arccosh^2(\sigma)+\Li_2\left[2-2\sigma(\sigma+\sqrt{\sigma^2-1})\right]\right)\right)+\mathcal
  O(\epsilon^0)\,.
\end{multline}
%
 \subsection{The $s+u$-channel contribution}
The contribution of the combined $s$ and $u$ channel is simplified thanks
to symmetrization, and reads
\begin{multline}
J_{s-u}:=J_s+J_u=\frac{|\underline q|^{2D-10}}{96\hbar^3}\int \frac{d^D l_1 d^D l_2}{(2\pi)^{2D}} \frac{1}{(l_1^2+i\varepsilon) (l_2^2+i\varepsilon) ((l_1+l_2+u_q)^2+i\varepsilon)}\cr\sum_{1 \leq 
i \ne j \leq 3} \sum_{1 \leq k \ne n \leq 3} \frac{1}{\displaystyle\Big(\bar{p}_1 \cdot l_i+|\vec{q}| \frac{u_q \cdot l_i}{2}+i \varepsilon\Big)\Big(\bar{p}_1 \cdot l_j-|\vec{q}| \frac{u_q \cdot l_j}{2}-i \varepsilon\Big)}\cr\times\frac{1}{\displaystyle\Big(\displaystyle\bar{p}_2 \cdot l_k-|\vec{q}| \frac{u_q \cdot l_k}{2}-i \varepsilon\Big)\Big(\displaystyle \bar{p}_2 \cdot l_n+|\vec{q}| \frac{u_q \cdot l_n}{2}+i \varepsilon\Big)}\,.
\end{multline}
We introduced the momenta $\bar{p_i}=p_1-q/2$ and $\bar p_2=p_2+q/2$
which are orthogonal to $q$. And as before we rescaled the loop
momenta by $\ell_i\to l_i |q|$ with $q=u_q |q|$,  and $q=\hbar \underline q$.\\[5pt]
In this expression we have neglected the $l_i^2$ terms in the
denominators in the sum for the reason that they do not contribute to the classical limit.  In the
$\hbar$ expansion the $l_i^2$ contribution will cancel a massless
propagator and give a reduced graph that does not have a classical
limit, in a similar way as we have seen when expanding eq.~\eqref{e:Ionedelta}.\\[5pt]
Because
the integral is of order $1/\hbar^3$ one needs to expand the
denominator in the integral to the second order in $\hbar^2$ for
getting the  classical contribution
\begin{equation}
    J_{s -u}= J_{s-u }^0+\hbar |\underline q| J_{s-u
    }^1+\hbar^2|\underline q|^2 J_{s -u}^2+\mathcal O(|\vec q|^3)\,.
  \end{equation}
  \subsubsection{Evaluation of $J_{s-u}^0$}
\label{sec:lead-order-contr}
The leading order expansion in $|q|$ of the double-box contribution
$J_{s-u}$ is given by 
\begin{multline}
J_{s-u }^0=\frac{ |\underline q|^{2D-10}}{96\hbar^3}\int \frac{d^D l_1 d^D l_2}{(2\pi)^{2D}} \cr\times\sum_{1 \leq i \ne j \leq 3}  \sum_{1 \leq 
k \ne n \leq 3} \frac{1}{(\bar{p}_1 \cdot l_i+i \varepsilon)(\bar{p}_1 \cdot 
l_j-i \varepsilon)(\bar{p}_2 \cdot l_k-i \varepsilon)(\bar{p}_2 \cdot l_n+i \varepsilon)l_1^2 l_2^2 (l_1+l_2+u_q)^2}\,,
\end{multline}
which can  be rewritten in terms of delta-functions as
\begin{equation}
J_{s -u}^0=\frac{|\underline q|^{2D-10}}{96\hbar^3} \int \frac{d^D l_1 d^D l_2}{(2\pi)^{2D-4}} \frac{\delta(\bar{p}_1 \cdot l_1)\delta(\bar{p}_1 
\cdot l_2)\delta(\bar{p}_2 \cdot l_1)\delta(\bar{p}_2 \cdot l_2)}{l_1^2 l_2^2 (l_1+l_2+u_q)^2}\,,
\end{equation}
and evaluated to 
\begin{equation}
J_{s-u }^0=-\frac{|\underline q|^{2D-10}}{96\hbar^3} \frac{1}{(\bar p_1\cdot
  \bar p_2)^2-\bar p_1^2\bar p_2^2} \int \frac{d^{D-2} \vec{l_1} d^{D-2} \vec{l_2}}{(2\pi)^{2D-4}} \frac{1}{\vec{l_1}^2 \vec{l_2}^2 (\vec{l_1}+\vec{l_2}+\vec{u_q})^2}\,,
\end{equation}
using that $(\bar p_1\cdot
  \bar p_2)^2-\bar p_1^2\bar p_2^2=m_1^2
  m_2^2(\sigma^2-1)-\frac{|\vec{q}|^2 s}{4}$ and the integrals in
  Appendix~A of~\cite{SmirnovEvaluating} give that 
\begin{equation}
J_{s-u }^0 =-\frac{|\underline q|^{2D-10}}{96\hbar^3}\frac{\Gamma(\frac{D}{2}-2)^3 \Gamma(5-D)}{(4 \pi)^{D-2} \Gamma(\frac{3(D-4)}{2})(m_1^2 m_2^2(\sigma^2-1)-\hbar^2\frac{|\underline q|^2 s}{4})}\,.
\end{equation}
Expanding the denominator  in $|\vec{q}|$ gives a contribution to the
classical order
\begin{multline}
J_{s-u }^0=-\frac{|\underline q|^{2D-10}}{96\hbar^3}\frac{\Gamma(\frac{D}{2}-2)^3
  \Gamma(5-D)}{(4 \pi)^{D-2} \Gamma(\frac{3(D-4)}{2})m_1^2
  m_2^2(\sigma^2-1)}\\
-\frac{|\underline q|^{2D-8}}{384\hbar}\frac{\Gamma(\frac{D}{2}-2)^3
  \Gamma(5-D)}{(4 \pi)^{D-2} \Gamma(\frac{3(D-4)}{2})m_1^4
  m_2^4(\sigma^2-1)^2}+ O(|q|^{2D-6})\,.
\end{multline}
The first term of order $1/\hbar^3$ is the super-classical
contribution,  whereas the next term of order $1/\hbar$ is a
classical contribution.
\begin{equation}\label{e:Jsu0result}
   J_{s-u }^0=-\frac{1}{384 |\underline q|^{4\epsilon +2} \hbar^3}\frac{\Gamma(-\epsilon)^3 \Gamma(1+2\epsilon)}{(4 \pi)^{2-2\epsilon} \Gamma(-3\epsilon)}\frac{4m_1^2 m_2^2(\sigma^2-1)+|\vec{q}|^2 s}{m_1^4 m_2^4(\sigma^2-1)^2}.
\end{equation}
%
\subsubsection{The evaluation of $J_{s-u}^1$}
\label{sec:sub-leading-order}
The sub-leading order expansion in $|q|$ of the double-box contribution
$J_{s-u}$ is given by 
\begin{multline}
\hbar |\underline q| J_{s -u}^1=-\frac{|\underline q|^{2D-9}}{96\hbar^2}\int
\frac{d^D l_1 d^D l_2}{(2\pi)^{2D-3}} \frac{3i (1-2u_q \cdot l_1)
  \delta(\bar{p}_1 \cdot l_1)\delta(\bar{p}_2 \cdot l_1)\delta(\bar{p}_2 \cdot l_2)}{
  2(\bar{p}_1 \cdot l_2)^2 l_1^2 l_2^2 (l_1+l_2+u_q)^2}\\
-\frac{|\underline q|^{2D-9}}{96\hbar^2}\int \frac{d^D l_1 d^D l_2}{(2\pi)^{2D-3}} \frac{3i (1-2u_q \cdot l_1) \delta(\bar{p}_1 \cdot l_1)\delta(\bar{p}_1 \cdot l_2)\delta(\bar{p}_2 \cdot l_1)}{2(\bar{p}_2 \cdot l_2)^2 l_1^2 l_2^2 (l_1+l_2+u_q)^2}\,.
\end{multline}
This integral is of order $|\underline q|^{2D-9}/\hbar^2$ and imaginary.  
The
$|q|^2$ expansion cannot lead to a contribution to the classical part
of the amplitude of order $|\underline q|^{2D-8}/\hbar$. So
this integral will not contribute to the classical result but to the super-classical one.\\[5pt]
The integral $ J_{s-u }^1$  is 
\begin{equation}\label{e:Jsu1result}
\hbar |\underline q|    J_{s-u }^1=\frac{i(m_1+m_2) |\underline q|^{-1-4\epsilon}}{512 \hbar^2\pi^{\frac52} m_1^3 m_2^3(\sigma^2-1)^{\frac32}}\frac{(4\pi)^{2\epsilon} \Gamma(\frac12-\epsilon)^2\Gamma(\frac12+2\epsilon)\Gamma(-\epsilon)\Gamma(\frac12-2\epsilon)}{\Gamma(\frac12-3\epsilon)\Gamma(-2\epsilon)}\,,
  \end{equation}
and will not contribute to the classical limit.

\subsubsection{The evaluation of $J_{s-u}^2$}
\label{sec:subs-sub-leading}
The contribution at the order $O(|\underline q|^2)$ is decomposed into several pieces (each one being an integral with two $\delta$ functions) that we treat in order
\begin{equation}
\hbar^2 |\underline q|^2   J_{s-u }^2=J_{\rhd}^2+J_{\Box}^{2,\perp}+J_{\lhd}^2+J_{\Box}^{2,\parallel}\,.  
\end{equation}
\noindent $\bullet$  $J_{\rhd}^2$  is expanded on the master integrals of 
section~\ref{sec:masterI} using {\tt LiteRed}
\begin{equation}
  J_{\rhd}^2={|\underline q|^{-4\epsilon}\over 64 m_1^2m_2^4
    (\sigma^2-1)^2\hbar}\lim_{\sigma\to1}\left({2\over3} {\mathcal
      I_2(\sigma)\over2\epsilon^4 \sqrt{\sigma^2-1}}- {3+4\epsilon\over3}
    {\mathcal I_9^{+-}(\sigma)\over \epsilon^4}\right)\,,
\end{equation}
using the results in section~\ref{sec:constants} we obtain
\begin{equation}\begin{split}
J_{\rhd}^2&=\frac{|\underline {q}|^{-4\epsilon}}{64 m_1^2 m_2^4 (\sigma^2-1)^2\hbar}\left(\frac{b_4}{3 \epsilon}-\frac{(3+4\epsilon)b_9^{+-}}{3\epsilon^2}\right)\cr &=
\left(\frac{4\pi e^{-\gamma_E}}{|\underline{q}|^2}\right)^{2\epsilon} \frac{1+\epsilon}{2048 \pi^2 \epsilon^2 m_1^2 m_2^4 (\sigma^2-1)^2\hbar}+\mathcal O(\epsilon^0)\,.
\end{split}\end{equation}
\noindent $\bullet$  The contribution $J_{\lhd}^2$ is obtained by exchanging $m_1$ and
$m_2$ and is given by 
\begin{equation}
J_{\lhd}^2=\left(\frac{4\pi e^{-\gamma_E}}{|\underline{q}|^2}\right)^{2\epsilon} \frac{1+\epsilon}{2048 \pi^2 \epsilon^2 m_1^4 m_2^2 (\sigma^2-1)^2\hbar}+\mathcal O(\epsilon^0)\,. %
\end{equation}
\noindent $\bullet$  The decomposition of $J_{\Box}^{2,\perp}$  on the master integrals
with {\tt LiteRed} is given by 
\begin{multline}
J_{\Box}^{2,\perp}=-\frac{|\underline q|^{-4\epsilon}}{64 m_1^3 m_2^3
  (\sigma^2-1)^2\hbar}\Big(-\frac{8\sqrt{\sigma^2-1}}{\epsilon^2} \mathcal
I_2(\sigma)+\frac{2(1+2\epsilon)\sqrt{\sigma^2-1}}{\epsilon^3} \mathcal
I_3(\sigma)\cr
-\frac{2(1+6\epsilon)\sigma}{3\epsilon^3} \mathcal I_4(\sigma)
-\frac{(3+2\epsilon)\sigma}{3\epsilon^4} (\mathcal I_9^{+-}(\sigma)+\mathcal I_9^{++}(\sigma))\Big)\,,
\end{multline}
using the result from the evaluation of the master integrals in
section~\ref{sec:masterI} we get
\begin{multline}
J_{\Box}^{2,\perp}=\frac{-|\underline q|^{-4\epsilon}}{64 m_1^3
  m_2^3 (\sigma^2-1)^2\hbar\epsilon^2}\Big(\frac{-(2\epsilon b_4\!+\!(3\!+\!2\epsilon)(b_9^{+-}\!+\!b_9^{++}))\sigma}{3}+2(1+2\epsilon)b_3(\sigma^2-1)^{\frac{1}{2}-\epsilon}\cr
-2\sigma b_3\int_1^{\sigma}{dt\over (t^2-1)^{\frac{1}{2}+\epsilon}}\Big)+\mathcal O(\epsilon^0)\,.
\end{multline}

\noindent $\bullet$  For $J_{\Box}^{2,\parallel}$  we have
\begin{multline}
J_{\Box}^{2,\parallel}=-\frac{|\underline
  q|^{-4\epsilon}}{256\hbar}\int \frac{d^D l_1 d^D l_2}{(2\pi)^{2D-2}}
\frac{(1-2u_q \cdot l_1)^2 \delta(\bar{p}_1 \cdot l_1)
  \delta(\bar{p}_2 \cdot l_1)}{l_1^2 l_2^2 (l_1+l_2+u_q)^2}\cr \times
\Big( \frac{1}{(\bar{p}_1\cdot l_2-i\varepsilon)^2(\bar{p}_2\cdot
  l_2+i\varepsilon)^2}+\frac{1}{(\bar{p}_1\cdot
  l_2+i\varepsilon)^2(\bar{p}_2\cdot l_2+i\varepsilon)^2}\Big)\,,
\end{multline}
which evaluates to
\begin{equation}
J_{\Box}^{2,\parallel}=\left(\frac{4\pi e^{-\gamma_E}}{|\underline q|^2}\right)^{2\epsilon}\frac{i (1+2\epsilon)(\sigma \arccosh(\sigma)-\sqrt{\sigma^2-1}-\frac{i\pi \sigma}{2})}{1024\epsilon^2 \pi^{3}m_1^3 m_2^3(\sigma^2-1)^{2}\hbar}+\mathcal O(\epsilon^0)\,.
\end{equation}
Summing all these contributions  and using that for $\epsilon<0 $ and $\sigma>1$\footnote{We recall that  $\arccosh(\sigma)= \log(\sigma+\sqrt{\sigma^2-1})$ for $\sigma>1$.
}
\begin{align}\label{e:Intepsilonsigma}
\int_1^\sigma{dt\over (t^2-1)^{\frac12+\epsilon}}&=4^\epsilon\int_1^{\sigma+\sqrt{\sigma^2-1}}
                                                  {
                                                   x^{-1+2\epsilon}\over
                                                   (x^2-1)^{2\epsilon}} dx\\
\nonumber &={1\over(\sigma^2-1)^\epsilon}\Bigg( \arccosh(\sigma)
  -\epsilon\Big(\arccosh(\sigma)^2\cr&\hskip3cm+\Li_2\left(2-2\sigma(\sigma+\sqrt{\sigma^2-1})\right)
\Big) +\mathcal O(\epsilon)\Bigg),
\end{align}
we have
\begin{multline}\label{e:Jsu2result}
\hbar^2 \underline q^2 J_{s-u }^2=\left(\frac{4\pi e^{-\gamma_E}}{|\underline
    q|^2}\right)^{2\epsilon} \frac{1}{2048 \epsilon^2 \pi^3 m_1^3
  m_2^3 (\sigma^2-1)^2\hbar }\cr
\times\Bigg( \frac{(1+\epsilon)\pi s}{m_1
  m_2}+2i (1+2\epsilon)(\sigma \arccosh(\sigma)-\sqrt{\sigma^2-1})\cr
+2i \left(\frac{-1}{4(\sigma^2-1)}\right)^{\epsilon} \bigg(\sigma
\Big(\arccosh(\sigma)-\epsilon \big(\arccosh^2(\sigma)+\Li_2
(2-2\sigma(\sigma+\sqrt{\sigma^2-1}))\big) \Big)\cr
-(1+2\epsilon)\sqrt{\sigma^2-1})\bigg) \Bigg)+\mathcal O(\epsilon^0)\,.
\end{multline}
%
\subsubsection{The final expression for $J_{s-u}$}
\label{sec:final-expression-j_s}
The sum of all the contributions is given by
\begin{multline}
J_{s-u }=\frac{1}{(4\pi)^4\hbar}\left(\frac{4\pi e^{-\gamma_E}}{|\underline
    q|^2}\right)^{2\epsilon} \bigg(-\frac{\pi^2}{2 \epsilon^2
  \hbar^2 |\underline q|^2m_1^2m_2^2(\sigma^2-1)}\cr+\frac{i \pi(1+2\epsilon)(\sigma
  \arccosh{\sigma}-\sqrt{\sigma^2-1})}{4\epsilon^2 m_1^3
  m_2^3(\sigma^2-1)^{2}}
+ \frac{\pi^2 s}{8 \epsilon m_1^4 m_2^4 (\sigma^2-1)^2} \cr
+\frac{i \pi}{4\epsilon^2m_1^3 m_2^3 (\sigma^2-1)^2}
\left(\frac{-1}{4(\sigma^2-1)}\right)^{\epsilon} \cr\times\bigg(\sigma
\Big(\arccosh(\sigma)-\epsilon \big(\arccosh^2(\sigma)+\Li_2
(2-2\sigma(\sigma+\sqrt{\sigma^2-1}))\big) \Big)
-(1+2\epsilon)\sqrt{\sigma^2-1})\bigg) \Bigg)\bigg)+\mathcal O(\epsilon^0),
\end{multline}
we recall that the $1/\hbar$ contribution is of order $\epsilon$ and
does not contribute to this order.

\section{Differential equations for the master-integrals}
\label{sec:masterI}
For the computation of the two-loop integrals we need to evaluate the following nine
master integrals
\begin{align}
\label{e:I1def}\mathcal I_1(\sigma)&=2\epsilon^4 \mathcal I^{0,0,1,1,1,1,0},\\
\label{e:I2def}\mathcal I_2(\sigma)&=2\epsilon^4 \sqrt{\sigma^2-1}\mathcal I^{0,0,0,0,1,1,1},\\
\label{e:I3def}\mathcal I_3(\sigma)&=2\epsilon^3 \sqrt{\sigma^2-1}\mathcal I^{0,0,0,0,1,1,2},\\
\label{e:I4def} \mathcal I_4(\sigma)&=4\epsilon^2 (\sigma^2-1)\mathcal I^{-1,-1,0,0,1,1,3}+\epsilon^2(1+2\epsilon) \sigma \mathcal I^{0,0,0,0,1,1,2},\\
\label{e:I5def} \mathcal I_5(\sigma)&=\frac{2\epsilon^2(4\epsilon-1)(2\epsilon-1)}{\sqrt{\sigma^2-1}}\mathcal I^{0,0,0,1,0,1,1},\\
\label{e:I6def} \mathcal I_6(\sigma)&=2\epsilon^4 \sqrt{\sigma^2-1}\mathcal I^{0,0,1,1,1,1,1},\\
\label{e:I7def}\mathcal I_{7}(\sigma)&=8\epsilon^4(\sigma^2-1)\mathcal I^{-1,-1,1,1,1,1,1}+4\epsilon^4 \sigma \mathcal I^{0,0,1,1,1,1,0},\\
\label{e:I8def} \mathcal I_8(\sigma)&=-\epsilon^3 \mathcal I^{0,1,0,0,1,1,2},\\
\label{e:I9def} \mathcal I_9(\sigma)&=\epsilon^4 \mathcal I^{1,1,0,0,1,1,1},
\end{align}
with the following definition for the master integral
\begin{multline}\label{e:Imasterdef}
\!\!\!\!\!\!\mathcal I^{n_1,n_2,n_3,n_4,n_5,n_6,n_7}\equiv\int
\frac{d^{D-1}l_1d^{D-1}l_2}{(2\pi)^{2D-2}} \frac{1}{(k \cdot
  l_1)^{n_1} (k \cdot l_2)^{n_2}(l_1^2)^{n_3}((u_q+l_2)^2)^{n_4}
  ((l_1-u_q)^2)^{n_5} }\cr
\times \frac{1}{ (l_2^2)^{n_6}((l_1+l_2)^2-2(\sigma-1) k \cdot l_1 k \cdot l_2)^{n_7}},
\end{multline}
where we have  defined $k^2\equiv u_q^2\equiv-1$ and $k\cdot u_q\equiv
0$.

In  appendix~\ref{sec:deltatomaster} we explain how to convert
the two-loop integrals in $D$ dimensions  with delta-function insertions in the numerator
to two-loop integrals with a generalized propagator in $D-1$ dimensions. \\[5pt]
Using {\tt LiteRed}~\cite{Lee:2013mka} we find the following differential
system of equations
\begin{equation}
\!\!\!\frac{d}{d \sigma}\! \!\begin{pmatrix}\! \mathcal I_1(\sigma)\! \\ \! \mathcal I_2(\sigma) \! \\ \!\mathcal I_3(\sigma) \! \\ \!\mathcal I_4(\sigma) \! \\ \!\mathcal I_5(\sigma)\! \\ \!\mathcal I_6(\sigma) \! \\ \!\mathcal I_7(\sigma)\! \\ \!\mathcal I_8(\sigma) \! \\ \!\mathcal I_9(\sigma)\! \end{pmatrix}\!=\!\epsilon \!\begin{pmatrix}  0 &0 &0 &0 &0 &0 &0 &0&0\\ 0 & \frac{6\sigma}{\sigma^2-1} & 0 & \frac{1}{\sqrt{\sigma^2-1}} & 0 & 0 & 0 &0&0\\ 0 & 0 & -\frac{2\sigma}{\sigma^2-1} & \frac{2}{\sqrt{\sigma^2-1}} & 0 & 0 & 0 &0&0 \\  0  & -\frac{12}{\sqrt{\sigma^2-1}} & -\frac{2}{\sqrt{\sigma^2-1}} &0& 0 & 0 & 0&0&0\\  0 &0 & 0 &0 & -\frac{2\sigma}{\sigma^2-1} &0&0 &0&0 \\ 0&0& -\frac{4\sigma}{\sigma^2-1} & -\frac{2}{\sqrt{\sigma^2-1}} & -\frac{4 \sigma}{\sigma^2-1}&-\frac{2\sigma}{\sigma^2-1} & \frac{2}{\sqrt{\sigma^2-1}}&0&0\\ - \frac{4}{\sigma^2-1}&- \frac{12}{\sqrt{\sigma^2-1}}&-\frac{8}{\sqrt{\sigma^2-1}}&0&- \frac{8}{\sqrt{\sigma^2-1}} &- \frac{2}{\sqrt{\sigma^2-1}} & \frac{2 \sigma}{\sigma^2-1} 
&0&0\\ 0 &0 & 0 &0 &0 &0 &0 &0&0\\ 0 &0 &\frac{1}{\sqrt{\sigma^2-1}}  &0 &0 &0 &0&0&0 \end{pmatrix}\!\! \begin{pmatrix} \!\mathcal I_1(\sigma)\!\\ 
\!\mathcal I_2(\sigma)\! \\ \! \mathcal I_3(\sigma)\! \\ \! \mathcal I_4(\sigma)\! \\ \! \mathcal I_5(\sigma)\! \\ \! \mathcal I_6(\sigma)\! \\ \! 
\mathcal I_7(\sigma)\! \\ \! \mathcal I_8(\sigma)\! \\ \! \mathcal I_9(\sigma)\!\end{pmatrix}.
 \end{equation}
For the resolution of this system of differential equation one must
pay attention to the fact that the $\epsilon\to0$ limit and
$\sigma\to1$ limit do not commute.
Solving the differential system in
an $\epsilon$ expansion for fixed $\sigma$ does not lead to the
correct answer at $\sigma=1$.\footnote{This phenomenon has already been
noticed in other contexts and it has been  illustrated  in section~7.3
of~\cite{Henn:2014qga} for instance.}
We solve the differential system using iterated integrals methods and
numerical methods and use  {\tt pySecDec}~\cite{Borowka:2017idc} for validating our analysis.
\subsection{The master integrals $\mathcal I_1(\sigma)$, $\mathcal
  I_5(\sigma)$ and $\mathcal I_8(\sigma)$}
The following integrals are easily integrated 
\begin{align}
\label{e:I1int} \mathcal I_1(\sigma)&=b_1 \epsilon^4,\\
  \label{e:I5int} \mathcal I_5(\sigma)&=b_5 \epsilon (\sigma^2-1)^{-\epsilon},\\
 \label{e:I8int} \mathcal I_8(\sigma)&=a_8,
\end{align}
where $b_1$, $b_5$ and $a_8$ are constants of integrations that will
be determined later. \\[5pt]
When deriving the expression for the master integrals we will keep the
factors  $ (\sigma^2-1)^{-\epsilon}$ explicit in order to keep a
control of the limits $\epsilon\to0$ and $\sigma\to1$. When solving this system of differential equation we assume that
$\epsilon<0$.
\subsection{The master integrals $\mathcal I_2(\sigma)$,  $\mathcal I_3(\sigma)$ and  $\mathcal I_4(\sigma)$ }
We remark that the equation for $\mathcal I_2(\sigma)$ 
\begin{equation}
\frac{d \mathcal I_2(\sigma)}{d \sigma}=\frac{6\epsilon \sigma}{\sigma^2-1}\mathcal I_2(\sigma)+ \frac{\epsilon}{\sqrt{\sigma^2-1}}\mathcal I_4(\sigma),
\end{equation}
leads to 
\begin{equation}
\mathcal I_2(\sigma)= a_2(\sigma^2-1)^{3\epsilon}+\epsilon
(\sigma^2-1)^{3\epsilon}  \int_1^\sigma 
{\mathcal I_4(t)\over (t^2-1)^{\frac{1}{2}+3\epsilon}} dt .
\end{equation}
Using the mean value theorem~\cite[\S~V.4]{Dieudonne} the integral can be 
written as
\begin{equation}
 (\sigma^2-1)^{3\epsilon}  \int_1^\sigma 
{\mathcal I_4(t)\over (t^2-1)^{\frac{1}{2}+3\epsilon}} dt=    (\sigma^2-1)^{1+3\epsilon}  
{\mathcal I_4(t(\sigma))\over (t(\sigma)^2-1)^{\frac{1}{2}+3\epsilon}},
\end{equation}
where $t(\sigma)\in[1,\sigma]$, and we deduce that the integral
vanishes in the limit $\sigma\to1$. For this one needs that $\mathcal I_4(\sigma)$
stays finite which is confirmed by an numerical evaluation with
  {\tt pySecDec}.\\[5pt]
We impose the regularity of  $\mathcal I_2(\sigma)$ at  $\sigma=1$,
which implies that  $a_2=0$ therefore 
\begin{equation}\label{e:I2int}
\mathcal I_2(\sigma)= \epsilon
(\sigma^2-1)^{3\epsilon}  \int_1^\sigma 
{\mathcal I_4(t)\over (t^2-1)^{\frac{1}{2}+3\epsilon}} dt .
\end{equation}
In a similar fashion we have
\begin{align}
\mathcal I_3(\sigma)&= b_3\epsilon(\sigma^2-1)^{-\epsilon}+2\epsilon
              (\sigma^2-1)^{-\epsilon}  \int_1^\sigma {\mathcal
              I_4(t)\over  (t^2-1)^{\frac{1}{2}-\epsilon}}dt,\cr
\mathcal I_4(\sigma)&= a_4-12\epsilon \int_1^\sigma {\mathcal I_2(t)\over
              (t^2-1)^{\frac{1}{2}} }dt-2\epsilon \int_1^\sigma
              {\mathcal I_3(t)\over (t^2-1)^{\frac{1}{2}}}dt.
\end{align}
This leads to the following $\epsilon$ expansion
\begin{multline}\label{e:I2int2}
\mathcal I_2(\sigma)= \Bigg(b_4(\sigma^2-1)^{3\epsilon} \int_{1}^{\sigma}
  {dt\over (t^2-1)^{\frac{1}{2}+3\epsilon}
  }\cr
  -2b_3(\sigma^2-1)^{3\epsilon}  \int_{1}^{\sigma}
 {dt_2\over  (t_2^2-1)^{\frac{1}{2}+3\epsilon}}\int_{1}^{t_2}{dt_1\over (t_1^2-1)^{\frac{1}{2}+\epsilon}} \Bigg)\epsilon^3  +\mathcal O(\epsilon^4),
\end{multline}
and
\begin{multline}\label{e:I3int2}
\mathcal I_3(\sigma)= {b_3 \epsilon\over (\sigma^2-1)^{\epsilon}}
  -\Big(4{b_3\over (\sigma^2-1)^{\epsilon}} \int_{1}^{\sigma}
 {dt_2 \over  (t_2^2-1)^{\frac{1}{2}-\epsilon} }\int_{1}^{t_2}
 {dt_1\over (t_1^2-1)^{\frac{1}{2}+\epsilon}}\cr
 -2 {b_4\over (\sigma^2-1)^{\epsilon} } \int_{1}^{\sigma} {dt\over (t^2-1)^{\frac{1}{2}-\epsilon}}\Big)\epsilon^3+\mathcal O(\epsilon^4),
\end{multline}
and
\begin{equation}\label{e:I4int2}
\mathcal I_4(\sigma)= \left(b_4-2 b_3 \int_{1}^{\sigma} {dt\over (t^2-1)^{\frac{1}{2}+\epsilon}}\right)\epsilon^2+\mathcal O(\epsilon^4).
\end{equation}
%

\subsection{The master integrals $\mathcal I_7(\sigma)$,  and  $\mathcal I_9(\sigma)$ }
Then $\mathcal I_9$ is obtained by
\begin{equation}\label{e:I9int}
\mathcal I_9(\sigma)=b_9\epsilon^2+\epsilon \int_1^\sigma \frac{\mathcal I_3(t)}{\sqrt{t^2-1}}dt
  =\left(b_9+ b_3 \int_{1}^{\sigma}{dt\over (t^2-1)^{\frac{1}{2}+\epsilon}}\right)\epsilon^2+\mathcal O(\epsilon^4).
\end{equation}
The integral $\mathcal I_7$ is given by
\begin{multline}
\mathcal I_7(\sigma)=a_7(\sigma^2-1)^{\epsilon}-2\epsilon
(\sigma^2-1)^{\epsilon}  \int_1^\sigma {6\mathcal I_2(t)+4\mathcal I_3(t)+4\mathcal I_5(t) +\mathcal I_6(t)\over
  (t^2-1)^{\frac{1}{2}+\epsilon}}dt.
\end{multline}
The regularity of the solution at $\sigma=1$ for $\epsilon<0$ imposes
that $a_7=0$.
Using the previous results and keeping only the terms up to $\mathcal O(\epsilon^2)$ gives
\begin{equation}\label{e:I7int}
\mathcal I_7(\sigma)=-8\epsilon^2 (b_3+b_5) (\sigma^2-1)^{\epsilon}  \int_1^\sigma{dt\over (t^2-1)^{\frac{1}{2}+2\epsilon}}+\mathcal O(\epsilon^3)\,.
\end{equation}
%

\subsection{The master integrals $\mathcal I_6(\sigma)$}
For $\mathcal I_6$ we have
\begin{equation}
\mathcal I_6(\sigma)={b_6\over(\sigma^2-1)^{\epsilon}}-{4\epsilon\over
  (\sigma^2-1)^{\epsilon}  }\int_1^\sigma{ t
 ( \mathcal I_3(t)+\mathcal I_5(t))\over  (t^2-1)^{1-\epsilon}} dt-{2\epsilon\over
(\sigma^2-1)^{\epsilon}}  \int_1^\sigma
{\mathcal I_4(t)-\mathcal I_7(t)\over  (t^2-1)^{\frac{1}{2}-\epsilon}}dt.
\end{equation}
Up to the order $\mathcal O(\epsilon)^4$ leads to 
\begin{multline}
\mathcal I_6(\sigma)={b_6\over(\sigma^2-1)^{\epsilon}}-{4\epsilon^2(b_3+b_5)\over
  (\sigma^2-1)^{\epsilon}  }\int_1^\sigma{ t
 dt\over  t^2-1}-{2\epsilon\over
(\sigma^2-1)^{\epsilon}}  \int_1^\sigma
{\mathcal I_4(t)-\mathcal I_7(t)\over  (t^2-1)^{\frac{1}{2}-\epsilon}}dt+\mathcal O(\epsilon)^4\,.
\end{multline}
The finiteness of this expression imposes that $b_5=-b_3$, so that
\begin{multline}\label{e:I6int}
  \mathcal I_6(\sigma)=\Big({b_6\over (\sigma^2-1)^{\epsilon}}
  -{2 b_4 \over (\sigma^2-1)^{\epsilon}  }\int_{1}^{\sigma}  {dt\over
  (t^2-1)^{\frac{1}{2}-\epsilon}}\cr
+{4 b_3\over (\sigma^2-1)^{\epsilon} } \int_{1}^{\sigma}{dt_2\over
  (t_2^2-1)^{\frac{1}{2}-\epsilon}}\int_{1}^{t_1} {dt_1\over (t_1^2-1)^{\frac{1}{2}+\epsilon}}\Big)\epsilon^3+\mathcal O(\epsilon^4).
\end{multline}
%

\subsection{Determination of the constants of integrations}\label{sec:constants}
We determine the constants of integration so that we have a solution
that is valid in all regimes and allows a control of the 
$\epsilon\to0$ and $\sigma\to1$ limits.

\subsubsection{$b_1$ constant}

Evaluating the $\mathcal I_1$ thanks to the results of the appendix~A
of~\cite{SmirnovEvaluating} we find

\begin{equation}
b_1=\frac{1}{2} \Big(\int \frac{d^{D-1} \vec{l}}{(2\pi)^{D-1}} \frac{1}{\vec{l}^2 (\vec{l}+\vec{u}_q)^2}\Big)^2=\frac{\Gamma(\frac{1}{2}+\epsilon)^2 \Gamma(\frac{1}{2}-\epsilon)^4}{2(4\pi)^{3-2\epsilon} \Gamma(1-2\epsilon)^2}.
\end{equation}

\subsubsection{$b_4$ constant}
For $b_4$ we have in $D=4-2\epsilon$
\begin{equation}\label{e:b4}
b_4=\lim_{\sigma \rightarrow 1} \frac{\mathcal I_2(\sigma)}{\epsilon^3\sqrt{\sigma^2-1}}=2 \epsilon \int \frac{d^{D-1} l_1 d^{D-1} l_2}{(2\pi)^{2D-2}}\frac{1}{l_1^2 l_2^2 (l_1+l_2+u_q)^2}=-\frac{2\epsilon \Gamma(\frac{1}{2}-\epsilon)^3\Gamma(2\epsilon)}{(4\pi)^{D-1}\Gamma(\frac{3}{2}-3\epsilon)}\,.
\end{equation}
So that at leading order in $\epsilon$,
\begin{equation}
b_4=-\frac{(4\pi e^{-\gamma_E})^{2\epsilon}}{32 \pi^2}+\mathcal O(\epsilon)\,.
\end{equation}

\subsubsection{$b_3$ and $b_5$ constants}\label{sec:b3b5}
The constants $b_3$ and $b_5=-b_3$ cannot be determined from the $\sigma=1$ limit, but we can do a direct computation of $\mathcal I_5$ and check the result with numerical analysis using {\tt
  pySecDec}~\cite{Borowka:2017idc}. We have using the definition of
$\mathcal I_5$\footnote{
  If one expands the master integral
  \begin{equation}
    \mathcal I^{0,0,0,1,0,1,1}= \int
    {d^{D-1}l_1d^{D-1}l_2\over(2\pi)^{2D-2} } {1\over (l_2+u_q)^2
      l_2^2 ((l_1+l_2)^2-2(\sigma-1)k\cdot l_1 k\cdot l_2)},
    \end{equation}
in  small $\sigma-1$ series and integrates terms by terms then the result is 0 by
dimension regularization of massless tadpoles. This is where we  see
the difference with the potential region computation in~\cite{Parra-Martinez:2020dzs} which has $\mathcal
I_5(\sigma)=0$. This would set $b_5=b_3=0$ in the computation and all
radiation reactions will cancel.}
  \begin{equation}
 \frac{\mathcal
   I_5(\sigma)}{2\epsilon^2(4\epsilon-1)(2\epsilon-1)}=\frac{\mathcal
   I^{0,0,0,1,0,1,1}}{\sqrt{\sigma^2-1}}=\frac{m_1 m_2}{\sqrt{\sigma^2-1}} \int \frac{d^D l_1
   d^D l_2}{(2\pi)^{2D-2}}
 \frac{\delta(\bar{p}_1\cdot l_1)\delta(\bar{p}_2\cdot l_2)}{l_2^2 (l_2-u_q)^2 (l_1+l_2)^2}.
  \end{equation}
  Then we write
  \begin{equation}
    \delta(\bar{p}_1 \cdot l_1)=\frac{1}{2i\pi}\left(\frac{1}{\bar{p}_1 
\cdot l_1-i\varepsilon}-\frac{1}{\bar{p}_1 \cdot l_1+i\varepsilon}\right),
  \end{equation}
  using the principal part identity in~\eqref{e:PP}
  and perform the $l_1$ integration using the results of the appendix~A
of~\cite{SmirnovEvaluating}, giving
 \begin{equation}
\frac{\mathcal I_5(\sigma)}{2\epsilon^2(4\epsilon-1)(2\epsilon-1)}=-\frac{4\Gamma(1-\epsilon)\Gamma(-1+2\epsilon)}{(4\pi)^{2-\epsilon}(\sigma^2-1)^{\epsilon}} \int \frac{d^{D-1} \vec{l_2}}{(2\pi)^{D-1}} \frac{(2 \vec{k} \cdot \vec{l_2})^{1-2\epsilon}}{\vec{l_2}^2 (\vec{l_2}-\vec{u_q})^2}.
  \end{equation}
Using again the results of the appendix~A of~\cite{SmirnovEvaluating} gives
\begin{equation}
\mathcal I_5(\sigma)=-\frac{i\epsilon  (4\pi)^{2\epsilon} }{32\pi^2  (\sigma^2-1)^{\epsilon}}\frac{(-1)^\epsilon \Gamma(1-2\epsilon)^2 
\Gamma(1+2\epsilon)^2 \Gamma(1-\epsilon)}{\Gamma(1-4\epsilon)\Gamma(1+\epsilon)},
\end{equation}
with $(-1)^\epsilon=e^{i\pi \epsilon}=1+i \pi \epsilon +\mathcal O(\epsilon^2)$. So at the end we get
\begin{equation}\label{e:b3b5}
b_5=-b_3=-\Big(-\frac{1}{4} \Big)^\epsilon \frac{i(4\pi e^{-\gamma_E})^{2\epsilon}}{32\pi^3}+\mathcal O(\epsilon^2).
\end{equation}
We have confirmed this result using numerical evaluations with {\tt pySecDec}~\cite{Borowka:2017idc}.

\subsubsection{$a_8$ constant}
For the constant $a_8$ we have to take care of the $i\varepsilon$ prescription, defining
\begin{equation}
a_8^{\pm}=-\epsilon^3 \int \frac{d^{D-1}l_1d^{D-1}l_2}{(2\pi)^{2D-2}} \frac{1}{(k \cdot l_2 \pm i \varepsilon)l_1^2 l_2^2 (l_1+l_2+u_q)^4}.
\end{equation}
Using the results of the appendix~A of~\cite{SmirnovEvaluating} gives
\begin{equation}
a_8^{\pm}=\pm \frac{i \sqrt{\pi} \epsilon^3 \Gamma(-\epsilon)\Gamma(-\frac{1}{2}-2\epsilon)\Gamma(\frac{3}{2}+2\epsilon)\Gamma(\frac{1}{2}-\epsilon)\Gamma(-\frac{1}{2}-\epsilon)}{(4\pi)^{3-2\epsilon}\Gamma(-2\epsilon)\Gamma(-\frac{1}{2}-3\epsilon)}.
\end{equation}
So that we get in the end
\begin{equation}
a_8^{\pm}=\mp \frac{i\varepsilon^3(4\pi e^{-\gamma_E})^{2\epsilon}}{32\pi}+\mathcal O(\epsilon^4).
\end{equation}

\subsubsection{$b_9$ constant}
Similarly for $b_9$ we have
\begin{equation}
b_9^{\pm \pm}=\epsilon^2\int \frac{d^{D-1}l_1d^{D-1}l_2}{(2\pi)^{2D-2}} 
\frac{1}{(k \cdot l_1\pm i \varepsilon) (k \cdot l_2 \pm i \varepsilon)l_1^2 l_2^2 (l_1+l_2+u_q)^2}.
\end{equation}
Using the results of the appendix~A of~\cite{SmirnovEvaluating} gives
\begin{equation}\label{e:b9}
b_9^{+-}=b_9^{-+}=-\frac{\epsilon^2}{6} \frac{\Gamma(-\epsilon)^3\Gamma(1+2\epsilon)}{(4\pi)^{2-2\epsilon} \Gamma(-3\epsilon)}=-\frac{(4\pi e^{-\gamma_E})^{2\epsilon}}{32 \pi^2}+\mathcal O(\epsilon^2),
\end{equation}
and
\begin{equation}
b_9^{++}=b_9^{--}=-2b_9^{+-}.
\end{equation}

%
\subsection{Final determination of the master integral $\mathcal I_6(\sigma)$}\label{e:I6evaluation}
Using that  the leading order expansion in $\epsilon\to0$ (with
$\epsilon<0$) and  $\sigma>1$ 
\begin{equation}
    \int_{1}^{\sigma}{dt\over
(t^2-1)^{\frac{1}{2}-\epsilon}}=\arccosh(\sigma)+\mathcal O(\epsilon),
\end{equation}
and 
\begin{equation}
  \int_{1}^{\sigma}{dt_2\over
  (t_2^2-1)^{\frac{1}{2}}}\int_{1}^{t_2} {dt_1\over
  (t_1^2-1)^{\frac{1}{2}}}={(\arccosh(\sigma))^2\over2}+\mathcal O(\epsilon),
\end{equation}
and using the values for the constant of integration $b_4$ in eq.~\eqref{e:b4} and
$b_3$  in eq.~\eqref{e:b3b5}  and that
$b_6=0$\footnote{This has been confirmed by numerical evaluations
  using {\tt pySecDec}~\cite{Borowka:2017idc}}, leads to 
\begin{equation}
\mathcal I_6(\sigma)=\frac{(4\pi e^{- \gamma_E})^{2\epsilon}
  \epsilon^3}{8\pi^3}
\arcsinh\left(\sqrt{\frac{\sigma-1}{2}}\right)\!\!\left(\pi\!+\!2i\left(-1\over  4(\sigma^2-1)\right)^{\epsilon}  \arcsinh\left(\sqrt{\frac{\sigma-1}{2}}\right)\!\!\right)+\mathcal O(\epsilon^4).
\end{equation}
%
\subsection{Final determination of the master integral $\mathcal I_9(\sigma)$}\label{sec:I9evaluation}
The master integral $\mathcal
I_9(\sigma)$ in eq.~\eqref{e:I9int}  reads
\begin{equation}
\mathcal I_9(\sigma)- b_9 \epsilon^2=b_3 \epsilon^2 \int_1^\sigma{dt\over (t^2-1)^{\frac12+\epsilon}} +\mathcal O(\epsilon^4)\,.
\end{equation}
The constant of integration $b_3$ has been determined
in eq.~\eqref{e:b3b5} and $b_9$  in eq.~\eqref{e:b9}.
Using the result in eq.~\eqref{e:Intepsilonsigma} for  the integral we have
\begin{equation}\label{e:I9res}\begin{split}
    \mathcal I_9 (\sigma)&= b_9 \epsilon^2+{b_3 \epsilon^2 \over (\sigma^2-1)^\epsilon}\Bigg( \arccosh(\sigma)
  -\epsilon\Big(\arccosh(\sigma)^2\cr & \hskip5cm+\Li_2\left(2-2\sigma(\sigma+\sqrt{\sigma^2-1})\right)
\Big) +\mathcal O(\epsilon)\Bigg)\,.
\end{split}\end{equation}
At high-energy $\sigma\gg1$  we have
\begin{equation}
\mathcal I_9(\sigma)\simeq b_9 \epsilon^2+{b_3 \epsilon^2\over(\sigma^2-1)^\epsilon}\left(
  \log(2\sigma)+\left(\zeta(2)+\log^2(2\sigma)\right)\epsilon\right)+\mathcal
                      O(\epsilon^4)\,.
 \end{equation}
%

\section{Conclusion}\label{sec:conclusion}
Motivated by the current interest in extending the computation of two-body interactions
in general relativity to the Post-Minkowskian regime, much effort has gone into establishing
a systematic approach based on modern scattering amplitude techniques. Critical to this
program is a complete understanding of how $S$-matrix calculations lead to classical scattering
information.\\[5pt]
We have here re-analyzed the evaluation of those integrals that contribute to classical observables
in gravity. We have aimed to simplify and streamline the known method of soft region evaluation
so as to make it amenable to high-order calculations. In the process of this program, we believe
we have clarified issues related to apparent differences, both between potential-region and
soft-region integrations
and between soft-region integrations that use different boundary conditions. At one-loop order,
our analysis confirms in a straightforward manner earlier results for both Einstein gravity and
${\cal N}=8$ supergravity. At two-loop order, we have focused exclusively on the simpler case of
${\cal N}=8$ supergravity. Our results here are in complete agreement with recent results
by Di Vecchia, Heissenberg, Russo, and Veneziano \cite{DiVecchia:2020ymx,DiVecchia:2021ndb,DiVecchia:2021bdo}. In the process of comparison,
we have made some observations that we hope will help clarify those computations.\\[5pt]
While our method described here is firmly anchored in the method of integration in the soft region,
it offers several advantages that will become further apparent at higher loop order. Most
importantly, by adding sets of integrands we recover combinations that precisely generate delta-functions,
and hence effectively reduce the dimensionality of integrations\footnote{Combining terms in this manner introduces delta-functions in an optimal way. This 
 generalizes the well-known one-loop eikonal procedure of~\cite{Kabat:1992tb}. The same idea was also used to simplify the $H$ and crossed-$H$ diagrams in~\cite{Parra-Martinez:2020dzs}.}. Besides, the combination
of diagrams before integrating leads to a quite
significant reduction in the number of master integrals that need to be evaluated.\\[5pt]
In addition to the real terms of the amplitude that will exponentiate in the eikonal limit, we have
also established in detail the relation between the real and imaginary part at two-loop order discussed in~\cite{DiVecchia:2021ndb,DiVecchia:2021bdo}. This seems to be a new situation where the cancellation of
infrared divergences must happen at the linear level in the $S$-matrix rather than in quantum mechanical
cross-sections. This issue should be better clarified in future work.
\subsection*{Acknowledgements}
We would like to thank Samuel Abreu,  Lorenzo
Magnea for discussions. We thank Roman Lee for having giving us access to 
the
version 2 of {\tt LiteRed} and answering questions about its usage. The research of P. V. has received funding from the ANR
grant ``Amplitudes'' ANR-17- CE31-0001-01, and the ANR grant ``SMAGP''
ANR-20-CE40-0026-01 and is partially supported by Laboratory of Mirror
Symmetry NRU HSE, RF Government grant, ag. No 14.641.31.0001. P.V. is
grateful to the I.H.E.S. for the use of their computer resources. The work of P.H.D. was supported in part by DFF grant 0135-00089A. The work of N.E.J.B.-B. was supported in part by the Carlsberg Foundation.
\newpage
\appendix
\section{Mapping to  integrals with a generalized propagator}\label{sec:deltatomaster}
In this appendix we explain how one can convert the two-loop integrals
with $\delta(\bar{p} \cdot l)$ insertions  (coming from the application of the
principle-part prescription in~\eqref{e:PP}) to integrals with a generalized
propagator $1/((l_1+l_2+u_q)^2+2(1-\sigma)(k \cdot l_1)(k \cdot l_2))$ that we use for the determination of the master integrals
in section~\ref{sec:masterI}.\\[5pt]
Using these manipulations we have the following correspondence
between the integrals. The dashed lines are massless propagators and
the wavy-lines are the generalized propagator.
\begin{align*}
   \includegraphics[width=7cm]{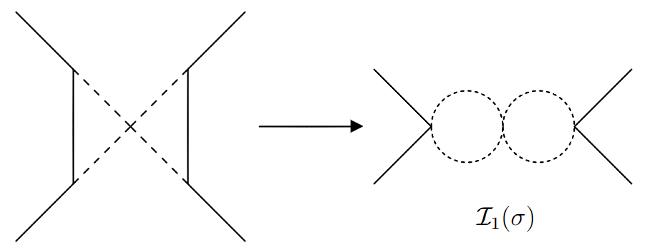}&
\includegraphics[width=7cm]{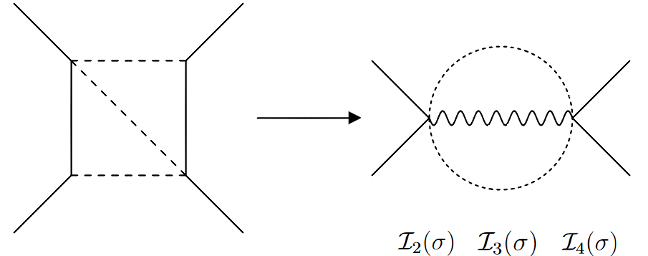}                          
\cr
  \includegraphics[width=7cm]{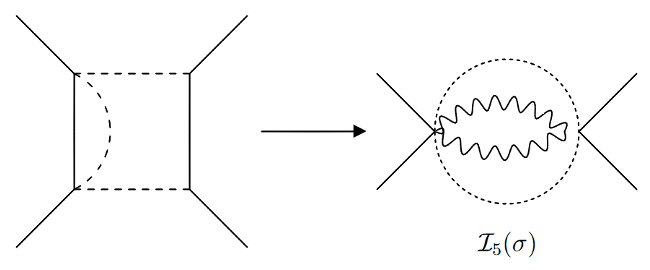}
&  \includegraphics[width=7cm]{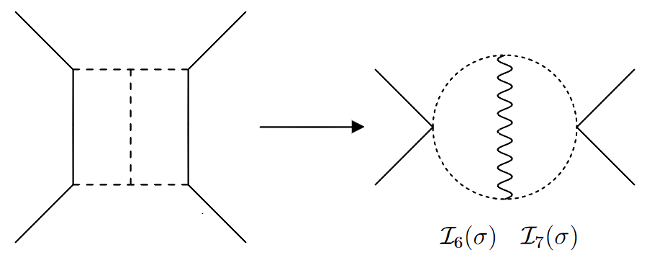}
 \cr
 \includegraphics[width=7cm]{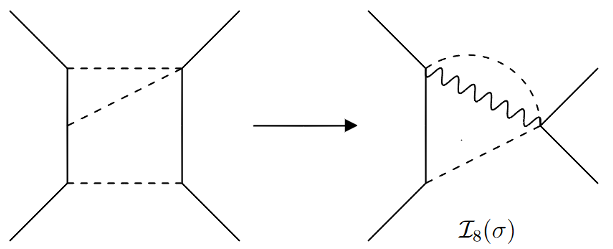}& \includegraphics[width=7cm]{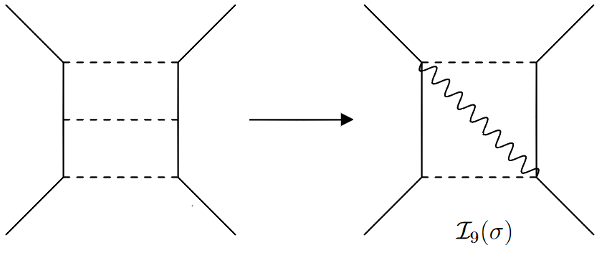}
 \nonumber
\end{align*}
We start with the two-loop $D$-dimensional  integral
\begin{equation}\begin{split}&
\mathcal D=\int \frac{d^{D}l_1d^{D}l_2}{(2\pi)^{2D-2}} \ \delta(\bar{p}_1 \cdot l_1)\delta(\bar{p}_2 \cdot l_2) \times  \\ & \frac{1}{(\bar{p}_1 
\cdot l_2\pm i \varepsilon)^{n_1} (\bar{p}_2 \cdot l_1\pm i \varepsilon)^{n_2}((u_q+l_1)^2)^{n_3}((u_q+l_2)^2)^{n_4} (l_1^2)^{n_5} (l_2^2)^{n_6}((l_1+l_2+u_q)^2)^{n_7}}.
\end{split}\end{equation}
We recall that $\bar p_1\cdot u_q=\bar p_2\cdot u_q=0$. Consider the
vector $k_1=\frac{-m_2 \sigma \bar p_1+m_1 \bar p_2}{m_1
  m_2\sqrt{\sigma^2-1}} $. The vector $k_1$ is in the plane generated
by $\bar p_1$ and $\bar p_2$, and we have $k_1^2=-1$ and $\bar p_1\cdot
k_1=0$. We have also $\bar p_2^{\mu}=\frac{m_2 \sigma}{m_1}\bar
p_1^{\mu}+m_2\sqrt{\sigma^2-1} k_1^{\mu}$ and $\bar p_2\cdot k_1=-m_2 \sqrt{\sigma^2-1}$.\\[5pt]
Separating the integration variables $l_i$ into
$l_i=\tilde{l_i}+\hat{l_i}$, where $\tilde{l_i}$ is the components of
$l_i$ inside the plane defined by $(p_1,p_2)$ and $\hat{l_i}$ are the
other components of the $D-2$ space, orthogonal to $(\bar p_1,\bar p_2)$.\\[5pt]
Note also that $(\frac{p_1}{m_1},k_1)$ forms an orthogonal basis of
the two dimensional space spanned by $(\bar p_1,\bar p_2)$. We write
$\tilde{l_2}=\frac{\bar p_1 \cdot l_2}{m_1}\frac{\bar p_1}{m_1} -(k_1
\cdot l_2)k_1=l_2^0 \frac{\bar p_1}{m_1}+l_2^1 k_1 $. Using the
Lorentz transformation
 \begin{equation}
 \begin{pmatrix} l_2^0  \\ l_2^1 \end{pmatrix}=\Lambda  \begin{pmatrix} 
L_2^0  \\ L_2^1 \end{pmatrix}= \begin{pmatrix} \sigma & \sqrt{\sigma^2-1} \\ \sqrt{\sigma^2-1} &\sigma \end{pmatrix} \begin{pmatrix} L_2^0  \\ L_2^1 \end{pmatrix},
\end{equation}
we have  $\bar p_2 \cdot l_2=m_2 L_2^0$ and $\bar p_1 \cdot
l_2=m_1\sigma L_2^0+m_1 \sqrt{\sigma^2-1} L_2^1$.
Similarly, we have $\bar p_1 \cdot l_1=m_1 l_1^0$, $\bar p_2 \cdot
l_1=m_2 \sigma l_1^0-m_2 \sqrt{\sigma^2-1} l_1^1$.\\[5pt]
We have as well that
\begin{equation}
    (l_1+l_2+u_q)^2=(l_1+L_2+u_q)^2+2(1- \sigma) (k_1 \cdot
    l_1)(k_1\cdot L_2),
\end{equation}
where we have used the conditions  $\bar p_1 \cdot l_1=\bar p_2 \cdot
l_2=0$ that  are imposed by the delta-functions.\\[5pt]
The implies that the integral with the delta-functions $\mathcal D$
can be written as 
\begin{multline}
\!\!\!\!\!\!\mathcal D=\frac{1}{m_1^{n_1+1}
  m_2^{n_2+1}(\sigma^2-1)^{\frac{n_1+n_2}{2}}}\int
\frac{d^{D}l_1d^{D}L_2}{(2\pi)^{2D-2}}\frac{\delta(m_1 l_1^0)\delta(m_2 L_2^0)}{(-k_1 \cdot l_2 \pm i \varepsilon)^{n_1} (k_1 \cdot l_1\pm i \varepsilon)^{n_2}((u_q+l_1)^2)^{n_3}} \cr \times\frac{1}{((u_q+L_2)^2)^{n_4} (l_1^2)^{n_5} (L_2^2)^{n_6}((l_1+L_2+u_q)^2+2(1-\sigma)(k_1 \cdot l_1)(k_1 \cdot L_2))^{n_7}},
\end{multline}
which leads to the $D-1$ integrals 
\begin{multline}
\!\!\!\!\!\!\mathcal D=\frac{(-1)^{n_1}}{m_1^{n_1+1} m_2^{n_2+1}(\sigma^2-1)^{\frac{n_1+n_2}{2}}}\int \frac{d^{D-1}l_1d^{D-1}l_2}{(2\pi)^{2D-2}} 
\frac{1}{(k \cdot l_2 \mp i \varepsilon)^{n_1} (k \cdot l_1\pm i \varepsilon)^{n_2}((u_q+l_1)^2)^{n_3}}\cr\times \frac{1}{((u_q+l_2)^2)^{n_4} (l_1^2)^{n_5} (l_2^2)^{n_6}((l_1+l_2+u_q)^2+2(1-\sigma)(k \cdot l_1)(k \cdot l_2))^{n_7}},
\end{multline}
which we use to build the basis of master integrals~\eqref{e:Imasterdef} at two-loop order.

\end{document}